\documentclass{article}

\usepackage{setspace}
\usepackage{shortcuts}
\usepackage{PRIMEarxiv}

\usepackage[utf8]{inputenc} 
\usepackage[T1]{fontenc}    
\usepackage{hyperref}       
\usepackage{url}            
\usepackage{booktabs}       
\usepackage{amsfonts}       
\usepackage{nicefrac}       
\usepackage{microtype}      
\usepackage{lipsum}
\usepackage{fancyhdr}       
\usepackage{graphicx}       
\graphicspath{{media/}}     

\usepackage[export]{adjustbox}
\usepackage{booktabs, threeparttable, dcolumn, multirow}
\usepackage{tabularx}
\usepackage{moresize}
\usepackage{etoolbox}
\AtBeginEnvironment{quote}{\par\singlespacing\small}
\usepackage{sankey}
\usepackage{rotating}
\usepackage{newfloat}
\usepackage{amsmath} 
\usepackage{algorithm}
\usepackage{algpseudocode}
\usepackage{natbib}
\usepackage{hyperref}
\hypersetup{
	colorlinks,
	linkcolor=red,
	citecolor=blue,
	urlcolor=blue
}

\usepackage{amssymb}

\title{What does the public want their local government to hear? \\A data-driven case study of public comments across the state of Michigan
}

\author{
  Chang Ge \\
  University of Michigan \\
  \texttt{changgge@umich.edu} \\
   \And
  Justine Zhang \\
  University of Michigan \\
  \texttt{tisjune@umich.edu} \\
  \And
  Haofei Xu \\
  Washington University in St. Louis\\
  \texttt{haofeix@wustl.edu} \\
  \AND
  Yanna Krupnikov \\
  University of Michigan \\
  \texttt{yanna@umich.edu} \\
  \And
  Jenna Bednar \\
  University of Michigan \\
  \texttt{jbednar@umich.edu} \\
  \And
  Sabina Tomkins \\
  University of Michigan \\
  \texttt{stomkins@umich.edu} \\
}

\begin{document}
\maketitle

\begin{abstract}
City council meetings are vital sites for civic participation where the public can speak directly to their local government. By addressing city officials and calling on them to take action, public commenters can potentially influence policy decisions spanning a broad range of concerns, from housing, to sustainability, to social justice. Yet studies of these meetings have often been limited by the availability of large-scale, geographically-diverse data. Relying on local governments' increasing use of YouTube and other technologies to archive their public meetings, we propose a framework that characterizes comments along two dimensions: the local concerns where concerns are situated (e.g., housing, election administration), and the \scs{} raised (e.g., functional democracy, anti-racism). Based on a large record of public comments we collect from 15 cities in Michigan, we produce data-driven taxonomies of the local concerns and \scs{} that these comments cover, and employ machine learning methods to scalably apply our taxonomies across the entire dataset. We then demonstrate how our framework allows us to examine the salient local concerns and \scs{} that arise in our data, as well as how these aspects interact.
\end{abstract}

\keywords{\textit{local government, public comments, political participation} \vspace{8ex}}

\section{Introduction}
In the United States, the public can, in theory, play a role in shaping local-level public policy \citep{tausanovitch2014representation,beierle2010democracy, holman2015, sahn2024public}.
Throughout the country, city council meetings typically contain public comment periods, where members of the public can speak directly to their local officials on a wide range of matters.
Participation in 
 local politics
has often been understood and encouraged as a means to address the unique, place-based concerns that arise in municipalities \citep{einstein2019local}.

Past research on public commenting in local politics
has  
studied activity around specific areas like housing \citep{einstein2019local,yoder2020a}
or school board administration \citep{adams2004public}, the representativeness of public commenters relative to
the general public 
\citep{mendelberg2000race,brady1995beyond, sahn2024public}, 
or the impacts of different forms of deliberation on civic engagement \citep{collins2021}. 
While rich, previous work tracking comments in city councils has often focused on particular issues (e.g. housing) and, at times, on individual municipalities (but see recent work from \citet{bararisimko2024} for a key exception).
As a result, it is often difficult to compare public commenting behavior across multiple issues and multiple places.

One reason that such comparisons are difficult is that studying city council meetings is difficult to scale. 
There are 19491 municipal governments in the United States \citep{uscb2022govorg}, and existing work has often depended on in person attendance of regular meetings or manual coding of large sets of transcripts. Since local governments perform many functions that can be the subject of public concern, there is a diverse and complex space of matters that commenters can bring up. This type of approach has meant ``extensive data collection and ... datasets that are necessarily limited in scope and static in time'' \citep[][p.224]{bararisimko2024}.

However, over the past decade, the growing reliance on digital media has created new opportunities to study public commentary during city council meetings. While open (``sunshine'') laws have long required public access to meeting minutes \citep{arnold2014}, an increasing number of municipalities have begun using technologies like YouTube and Granicus to fulfill these requirements by archiving full videos of meetings \citep{barari2023localview}. This trend accelerated during the COVID-19 pandemic, when many meetings transitioned to virtual formats \citep{einsteinetal2023}. The widespread adoption of YouTube has effectively standardized the recording of these meetings, as video postings limit the extent to which local governments can curate the format and content of meeting records \citep{bararisimko2024}.


Leveraging this increasing reliance on YouTube to archive local meetings, we propose a computational framework for characterizing public comments, deliberately designed to account for the wide range of concerns they cover. 
This framework addresses several gaps/limitations in the literature. 
Rather than preselecting topics with policy relevance like housing and school boards, 
our framework is amenable to a wide variety of policy concerns. 
Also, rather than relying on manual data annotation (hand coding)
we utilize a computational approach to tag topics at scale, across municipalities. 
Prior work has released a resource of YouTube channels 
of local governments \citep{barari2023localview}, which is useful in building a foundational entry-point 
into studying local government from digital media by offering metadata and limited text transcripts of meetings. 
Their goal is to offer access to some kinds of data about local government \citep{barari2023localview}. 
Our framework provides the ability to make sense of unstructured records 
by identifying specific local and societal concerns within discourse. 
Notably, we differentiate comments made by the public from those made by members of government 
and provide a machine learning approach for categorizing local and societal concerns. 
Our approach offers the opportunity to understand a much broader range of topics 
and at a larger scale than previous work.

In particular, our approach characterizes comments across two dimensions. The first dimension describes \textit{local concerns}, and draws on literature addressing the unique nature of local government. The second dimension draws on literature related to computational social science studies around issues 
from anti-racism to sustainability, and considers the extent to which local concerns are framed with broader \textit{\scs{}}. 
By connecting these two areas in a single study of public comments, we can track whether the people who comment engage both around 
local, place-based concerns and broader \scs{}. We are also able to explore how these two dimensions relate to each other. 
Moreover, we bridge several distinct bodies of literature by utilizing digital media to create this new framework. 

\section{Background}
Our approach draws on two themes of work. 
The first is the study of participation in local politics. 
The second is the study of political participation on social media, generally around large societal issues and social causes. 

\subsection{Background on participation in local politics}
Political participation has long been considered fundamental to a functional democracy, and participation in local politics has often been viewed as an especially important way for people to engage in the political processes that directly affect their lives \citep{verbaandnie1972, verba1995, verbaetal1993, eintsteinetal2019, mcleod1999}. Most municipalities across the United States regularly schedule time during council meetings for residents to directly address government officials. In this way, local governments are often the sites of peoples' most direct encounters with government processes and could, in theory, be gateways to greater engagement with democracy \citep{collins2021}.

Research in this space 
has typically 
centered on three themes:
(1) the structure and form of public deliberation
(2) the representativeness of commenters, and, importantly,
(3) the types of topics discussed. 
Although our work is specifically concerned with this last area -- the topical content of the public comments that people make at city council meetings -- we begin by considering the other themes as well.

\xhdr{The form of deliberation}
At their best, public comment periods can be considered a form of public deliberation \citep{collins2021}. To this end, research has considered whether different deliberative approaches could shift people's perceptions of and participation in local government meetings \citep{carpini2004public,collins2021,karpowitzmendelberg2014},
 as well as other modes of public participation such as protests \citep{novick2022black,hoang2024defund}. In turn, the extent to which local governments demonstrate a ``commitment to deliberative democracy,'' for example, can also shape the way local policymakers think about the broader concerns of the public \citep{collins2018}.
Although outside the scope of our goals in this paper, we note that our approach to considering public comments could be contextualized with research delving more deeply into the differences in institutional structures of comment periods. 

\xhdr{Representativeness in Participation}
Prior scholarship has theorized about inequitably distributed barriers to public participation \citep{brady1995beyond},
and has
demonstrated how public comments can reproduce social inequalities, with implications for eventual policy outcomes \citep{sahn2024public,einstein2019local, yoder2020a, collins2021, schaffneretal2020}.
Work on housing policy has, for example,
found stark demographic differences between local residents at large and those who opt to comment \citep{sahn2024public,einstein2019local}. Research also points to similar demographic differences between pro- and anti-development commenters,
 demonstrating that the share of preferences voiced by unrepresentative participants can predict whether housing projects are ultimately approved \citep [see][for review]{Brouwertrounstine2024}.
While our work does not consider commenters' demographic backgrounds, the content-centric framework we propose could scaffold future efforts that more richly characterize the ways in which different participants contribute in public meetings.

\xhdr{Topic Areas}
Research has long suggested that local governments are unique in the types of issues they address \citep{oliveretal2012, anzia2019}. From this perspective, many studies have long examined public comments on two distinctly local issues. One has been housing policy \citep{einstein2019local,yoder2020a,sahn2024public, tracy2007speaking,castro2022all,jenkins2023schools},
focusing on how participant comments on proposed zoning or development decisions.
Another focal area has been public schools, relying on public comment discourse in school board meetings \citep{tracy2007speaking,castro2022all,jenkins2023schools}. Jointly, this work has been foundational in developing an understanding of the way participation in public meetings can shape the course of local policy in these particular issue areas \citep{sahn2024public, yoder2020a}.  

Where work has considered issues more broadly, research has often focused on the \emph{nationalization} of local preferences, focusing on the preferences and comments of both, local policy-makers and ordinary voters. This work has tracked the extent to which local public opinion\citep[e.g.][]{hopkins2018increasingly} and local city council meetings \citep[e.g.][]{bararisimko2024} have increasingly focused on national issues. Here, research has also considered whether local issue positions are organized along similar dimensions as national issue positions \citep{schaffneretal2024}. 

Building on this research, our work aims to characterize the \textit{range} and \textit{relative prevalence} of issues people bring up across a broader set of cities, thus enabling future large-scale comparative analyses across multiple issues and places. 
We also expand beyond local issues, and consider the types of societal concerns raised in different places. 
This analysis is influenced by and contributes to the study of political participation on social media, which we discuss next. 

\subsection{Background on participation on social media around social causes} 
As people  use social media to engage politically,
\citep{bode2017gateway,mano2014social,jost2018social,bennett2012personalization,hosseinmardi2021examining}, there has been a wealth of work utilizing this digital trace data to understand public opinion around different issues. 
For example, \citeauthor{chang2023roeoverturned} collect data on Twitter 
to understand public reactions to the overturning of Roe v. Wade, \citep{chang2023roeoverturned}. 
Also exploring Twitter, \citeauthor{de2016social}, 
analyzed how participants on the platform participated in the Black Lives Matter movement \citep{de2016social}. 
Others have inspected public reactions to school shootings \citep{starbird2017examining}, 
climate change \citep{cody2015climate,dahal2019topic} and the COVID-19 pandemic \citep{lu2021public,dambanemuya2021characterizing,hsia2022characterizing}.

This work is deeply valuable in illuminating how people
react to different \scs{} on social media. They can offer broad pulses of how 
some participants feel about different issues and politics  \citep{nutakki2025there}. 
By mining public data, this work can provide insights into the community structure, 
media content, and the dynamics of political interest and participation \citep{jost2018social,jiang2023retweet,dubey2025talking,cascini2022social,lukito2021resonant,shugars2021pandemics}. 

However, we study the setting where the public speaks directly to local officials within a particular context. 
Thus, there is reason to think this sphere of participation will be different from social media. 
Furthermore, by studying this unique digital media setting, we avoid open challenges 
with using social media data to study political settings \citep{pearce2019social,tufekci2014big}.

\section{Our descriptive framework}
Using previous work as a foundation, we build our framework in a data-driven way.
Utilizing YouTube, we collect a dataset of public comments, drawing on the increasing availability of video recordings of council meetings online. 
By manually annotating a subset of this data, we develop taxonomies of the types of local concerns and \scs{} that commenters cover. 
Finally, we use machine learning models to apply these taxonomies and analyze the broader dataset. 

In sum, we contribute a computational framework consisting of
(1) a pipeline for collecting and extracting
records of public comments, 
(2) a data-driven taxonomy for describing local concerns
and \scs{}, 
and (3) a machine learning approach for applying
the taxonomy at scale.
Our framework enables us to answer the following research questions:

\begin{itemize}
    \item Which local concerns do people raise in their comments to their city councils? 
    \item Which \scs{} do people raise in their comments to their city councils? 
    \item How do local concerns and \scs{} interact? 
\end{itemize}

We illustrate how our framework is implemented and applied with a case study of public commenting activity 
across 15 cities in Michigan, USA.  While our focus is on Michigan, our approach scales to be applied much more broadly.
We begin by discussing the relevant literature which has inspired this taxonomy, then we describe how we operationalize 
local and societal concerns with our data. 

\section{A Taxonomy of Public Comments}
Focusing on the content of public comments, we aim to produce an account of public commenting that's rooted in how people engage with the specific functions of local government \citep{anzia2019, anzia2021}, while acknowledging that peoples' concerns are not always neatly scoped to local-level activities \citep[e.g.][]{hopkins2018increasingly}. Indeed, a cursory examination of our data suggests that many people use these venues to voice concerns that extend well beyond their locality, like sustainability, racial justice, and poverty.

Our initial manual annotations of the data, reinforced with previous research\cite{mayoralSurvey, local_governments_us, Avery2016, Einstein2017}, led us to two dimensions, which we preview prior to turning to the details of the method: the local concerns, and the \sc{}. Characterizing the public comments in terms of these two dimensions allows us to bridge the unique issue-focus of local government with the potential public interest in broader topics.

\xhdr{Local Concerns}
A comment's \emph{local concern} indicates the \textbf{actual local governance activities---e.g., policymaking or providing municipal services---the comment references}. Research has distinguished  topics which are ``uniquely local'' -- areas which are perceived to be the primary responsibility of local governments \citep{tausanovitch2014local}. Indeed, past work has reinforced local meetings as a space for these types of ``neighborhood concerns'' \citep{einstein2019local}. Here, we consider local concerns as these types of areas, which reflect people's ``place-based'' experiences \citep[e.g.][]{cramertoff2017, nuamah2021}, and could (reasonably) be addressed by local government. Note, however, commenters may raise concerns which do not neatly align with specific government roles and responsibilities -- but still emphasize a problem that is specific to the locality. Our set of local concerns reflects similar lists which have been used to distinguish ``distinctively'' local issues \citep{schaffneretal2024} but is grounded in the actual comments we study. 

\xhdr{Societal Concerns}
We define a \sc{} as a \textbf{broad concern regarding the wellbeing of individuals and groups}. These types of causes typically focus on broad areas, rather than having a direct connection to the function of the government. 
For example, we would define comments about the importance of sustainability generally as \scs{}; meanwhile, requesting a new recycling service in a park would be a local concern. Additionally, we  consider comments which speak out against racism, that is those which specifically mention that anti-racism is a value above and beyond specific policy implementations or local actions, as  raising
a societal concern.

Our approach to \sc{} is related to a body of work that uses digital trace data 
to understand public opinion around particular \scs{} through participation on social media (see \textbf{\textit{Background on participation on social media around social causes}}).
Instead of looking at public statements about politics in de-localized settings, 
we connect \scs{} to specific geographic places. 
In these settings, individuals may rarely bring up \scs{} at all, instead preferring to concentrate on specific local concerns that their officials can act on.

\xhdr{Local Concerns Intersecting with Societal Concerns}
By distinguishing these two dimensions we can examine the frames evoked when people articulate their thoughts on a policy; indeed, previous studies of housing \cite{einstein2019local} illustrate how people can voice opposition to development projects (e.g. local concern) in economic or aesthetic (e.g. \sc{}) terms.
Conversely, we can examine the dynamic ways that broad concerns are mapped onto actionable local outcomes; indeed, the Black Lives Matter protests in 2020 often concentrated broader racial justice demands (e.g. \sc{}) in calls for reforming or abolishing local police departments (e.g. local concern).
Crucially, having both dimensions allows us to account for the relationship between local issues and broader social concerns, enabling richer analyses.

Methodologically, our work adds to a large body of existing computational analyses of political processes \citep{grimmer2013text}.
While such studies have conventionally focused on national-level bodies such as the US Congress \citep{monroe2008fightin,quinn2010analyze,card2022immigration,gentzkow2019measuring}
or on mining public opinion around specific issues on social media \citep{starbird2017examining,jiang2023retweet,de2016social,chang2023roeoverturned},
a growing number of efforts have sought to compile datasets of local government proceedings, 
making use of governments' increasingly  common practice of providing online video recordings of their meetings \citep{brown2021council,barari2023localview,searchminutes}.
We extend these efforts by introducing a structured analytic framework focusing on public comments.

\section{A computational operationalization of our framework}
We
combine a data-driven approach with a novel framework for describing public comments. 
We first describe our data collection method, 
then our data-driven taxonomy of local concerns 
and \scs{}, 
and finally discuss the computational approach 
for applying this taxonomy to our data. 

\subsection{Data-collection pipeline}
To make use of the abundant data on public participation in city council meetings, 
we construct a pipeline for 
collecting and processing open records of city council meetings. 

To develop our approach and collect data in a tractable time frame, we start with a smaller set of cities in Michigan; we note that our framework is easily extensible to larger datasets, which we will collect in future work.
To select an informative and representative subset of Michigan cities, we follow an algorithmic procedure that selects a collection of cities whose distributions in terms of population and political leaning 
most closely match Michigan's statewide distributions (see Appendix Algorithm \ref{alg:goldset} for details).

\begin{figure*}[!htb]
    \includegraphics[width=\columnwidth]{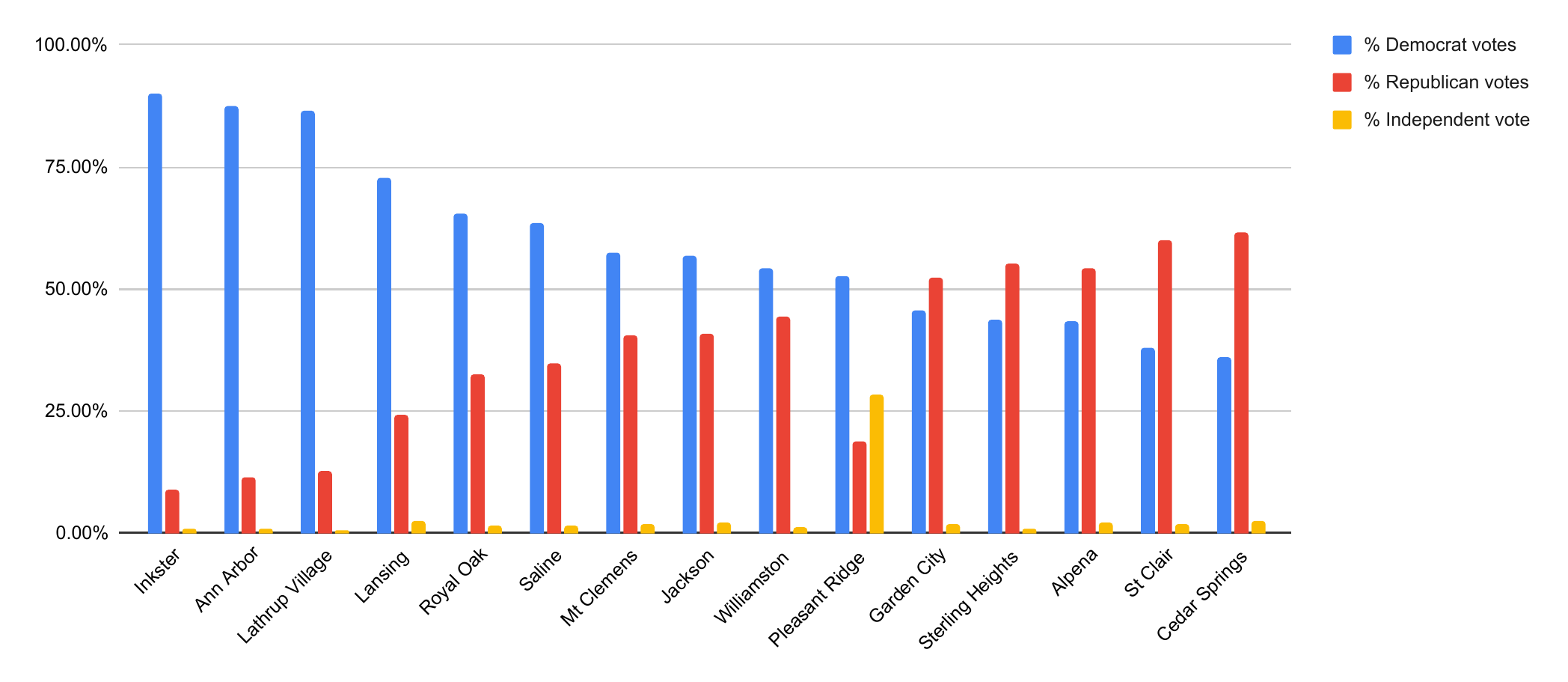}
    \caption{Partisan distribution of votes in the 2020 Presidential Election across selected Michigan cities. Showing the percentage of Democrat, Republican, and Independent votes in each city.}
    \label{fig:election_votes}
\end{figure*}

In \tabref{tab:desc_city}, we describe our dataset, which includes both large and small Michigan cities spanning a diversity of political leanings and socio-demographic characteristics.
For example, in Table \ref{tab:facts_city}, we can see that the smallest population in our dataset is 2610 residents (Pleasant Ridge), the median is 15735 (Mount Clemens), the maximum is 134062 (Sterling Heights); for reference, the statewide and national means are 9588 and 10615 respectively, for legally incorporated municipalities. 
For more information about the selected cities, please see Appendix Table \ref{tab:cityFacts}, Figure \ref{fig:population}, and Figure \ref{fig:election_votes}.

\begin{figure}[!htb]
    \includegraphics[width=\columnwidth]{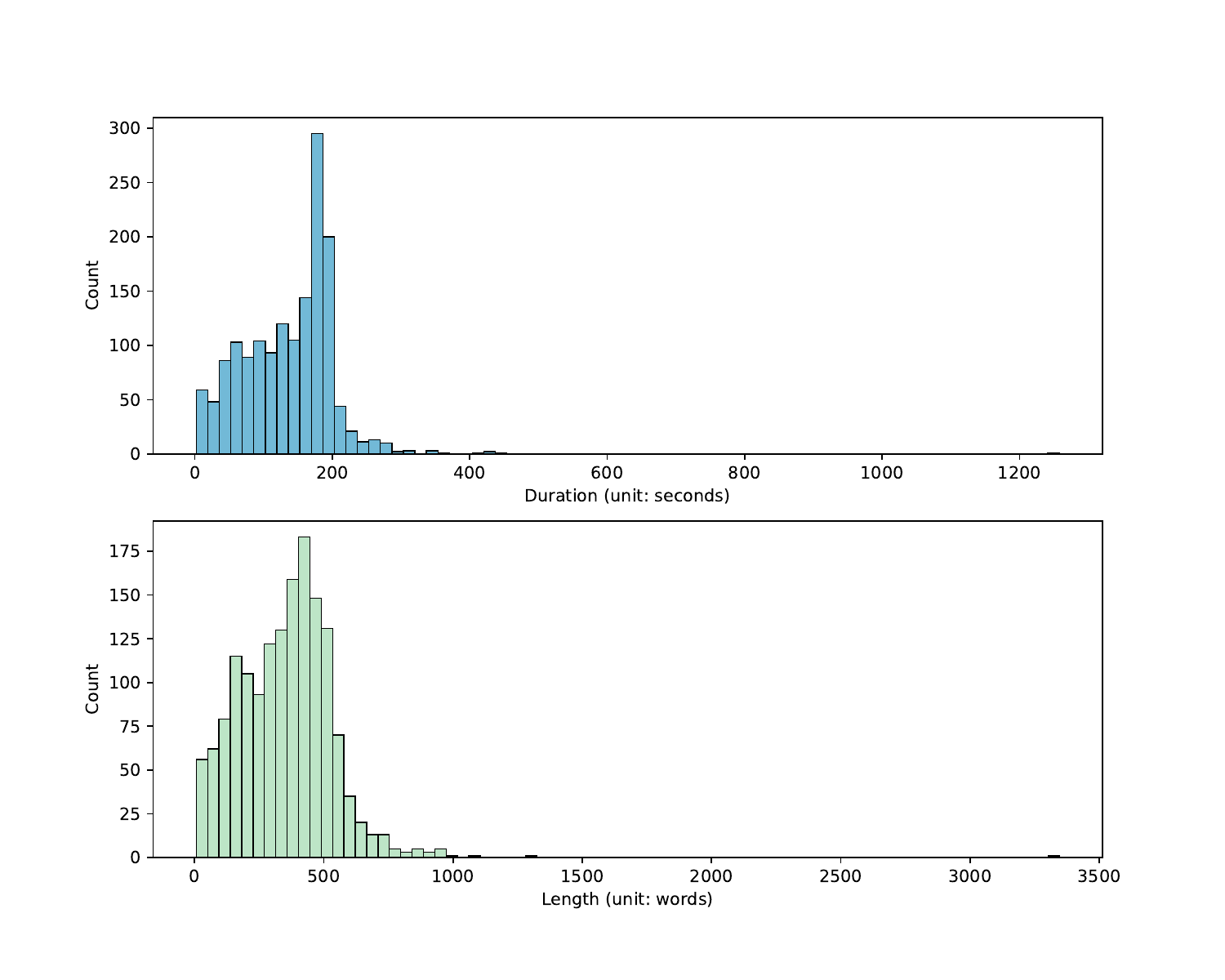}
    \caption{Comments distribution. \footnotesize The figure shows the histogram of comment lengths and comment durations. Most comments are relatively short, with an average duration of 136.43 seconds (2 minutes and 16.42 seconds) and an average length of 350.95 words. Most comments are short in both length and duration, likely due to city-imposed time limits for each commenter. For instance, Ann Arbor enforces a 3-minute limit, while Alpena allows up to 5 minutes. Outliers are present, with the longest comment (a comment from Cedar Springs) exceeding 20 minutes in duration and 3345 words in length. This is because while most cities do not respond to the comment immediately after, some smaller cities allow back and forth communication between council members and commenters during the public comment session.}
    \label{fig:duration_length}
\end{figure}

For each city, we collect and generate transcripts for meetings in 2023 from January 3 to December 19, using WhisperX \citep{bain2022whisperx}. 
First, we employ Whisper’s Voice Activity Detection (VAD) model \citep{radford2022robust} to segment the audio files.
These segments are then merged into clips of roughly 30 seconds transcribed by Whisper's Automatic Speech Recognition (ASR) model. Finally, we assign unique speaker IDs to each segment using the pyannote Speaker Diarization model \citep{Plaquet2023}.
Given the transcripts, we then extract the public comments made over the course of each meeting. For each meeting, we determine for each segment: whether it was an utterance during the public comment session, and the role of the speaker (government official or public commenter) of the utterance. We then focus on utterances by public commenters during the public comment session. 
After the initial data collection, we removed comments with fewer than 5 words in length or less than 2 seconds in duration and eliminated utterances that did not represent public comments
(e.g. ``Thanks'', ``Bye'', ``Can you hear me? (on Zoom)''). This process results in a dataset of 15 cities in the state of Michigan, 
259 meetings, and 1559 comments, see
\tabref{tab:desc_city}. For the distribution of duration and length of the comments, see \figref{fig:duration_length}.

\begin{table}[hbt!]
    \begin{adjustbox}{width=0.8\columnwidth,center}{
\def\sym#1{\ifmmode^{#1}\else\(^{#1}\)\fi}
\begin{tabular}{rccc}
\toprule
 & (1) & (2) & (3) \\
City & Meetings Count  &  Comments Count & Annotated Comments \\
\midrule
Ann Arbor         & 24 & 328 & 257 \\
Alpena        & 11 & 19 & 19  \\
Cedar Springs         & 11 & 51 & 51  \\
Garden City         & 21 & 47 & 47  \\
Inkster        & 20 & 148 & 148 \\
Jackson         & 20 & 198 & 198 \\
Lansing         & 26 & 249 & 193\\
Lathrup Village         & 9  & 56 & 0\\
Mount Clemens         & 22 & 54  &0\\
Pleasant Ridge         & 10 & 20  &0\\
Royal Oak         & 16 & 197 &0\\
Saint Clair         & 17 & 39  &0\\
Sterling Heights         & 19 & 74  &0\\
Saline         & 21 & 54  &0\\
Williamston        & 12 & 25  &0\\
\midrule
Total & 259 & 1559 & 913 \\ 
\bottomrule
\end{tabular}
}
\end{adjustbox}
    \caption{Column (1) shows the number of city council meetings we scraped for each city in 2023. Column (2) shows the total number of public comments from the respective city in 2023. Column (3) shows the number of comments in Column (2) that we annotated. If the number in Column (3) is greater than zero, we use comments from the respective city for model training (we call this annotated data in Figure \ref{fig:LocalGovKFold}); otherwise, we use comments from the respective city for label prediction (we call this unannotated data in Figure \ref{fig:LocalGovKFold}). }
    \label{tab:desc_city}
\end{table}

\begin{table}[hbt!]
    \begin{adjustbox}{width=\columnwidth,center}{
\def\sym#1{\ifmmode^{#1}\else\(^{#1}\)\fi}
\Huge
\begin{tabular}{lccccccc}
\toprule
       & (1) & (2)  & (3) & (4)  & (5) & (6) & (7) \\
       &  & Percentage of  & Percentage of & Median  & Percentage of  & Percentage of & Per capita  \\
   City    & Population &  Democrat votes & Republican votes & age  & White population &  Black population & income  \\
\midrule
Ann Arbor        & 122,731            & 87.64\%                      & 11.34\%                        & 25.9       & 68\%                           & 6\%                            & \$50,788            \\
Alpena           & 10,181             & 43.37\%                      & 54.44\%                        & 44.7       & 93\%                           & 2\%                            & \$30,936            \\
Cedar Springs    & 3,635              & 35.93\%                      & 61.56\%                        & 41.8       & 88\%                           & 0\%                            & \$25,042            \\
Garden City      & 27,268             & 45.79\%                      & 52.45\%                        & 40.9       & 85\%                           & 6\%                            & \$31,043            \\
Inkster          & 25,849             & 89.95\%                      & 8.99\%                         & 32.3       & 16\%                           & 75\%                           & \$22,420            \\
Jackson          & 31,810             & 56.88\%                      & 40.89\%                        & 35.0         & 65\%                           & 20\%                           & \$23,447            \\
Lansing          & 113,592            & 72.89\%                      & 24.41\%                        & 33.9       & 52\%                           & 22\%                           & \$29,023            \\
Lathrup Village  & 4,098              & 86.52\%                      & 12.74\%                        & 48.2       & 30\%                           & 64\%                           & \$54,461            \\
Mt Clemens       & 15,735             & 57.49\%                      & 40.65\%                        & 40.2       & 65\%                           & 24\%                           & \$31,939            \\
Pleasant Ridge   & 2,610              & 52.80\%                      & 18.89\%                        & 43.9       & 91\%                           & 1\%                            & \$84,370            \\
Royal Oak        & 58,368             & 65.59\%                      & 32.72\%                        & 35.9       & 83\%                           & 5\%                            & \$59,760            \\
St Clair         & 5,489              & 38.05\%                      & 60.02\%                        & 41.9       & 95\%                           & 2\%                            & \$44,068            \\
Sterling Heights & 134,062            & 43.83\%                      & 55.09\%                        & 40.7       & 80\%                           & 6\%                            & \$35,742            \\
Saline           & 9,072              & 63.46\%                      & 34.82\%                        & 44.6       & 86\%                           & 2\%                            & \$44,741            \\
Williamston      & 3,845              & 54.21\%                      & 44.41\%                        & 40.2       & 97\%                           & 0\%                            & \$51,635            \\ \midrule
\textbf{Mean across 15 cities}
                  & 37,890  & 59.63\% & 36.90\%   & 39.3\   & 73\%   & 16\%    & \$41,294    \\
\textbf{Michigan} 
                  & 9,588    & 50.62\%   & 47.84\%    & 40.9    & 87\%   & 5\%    & \$30,709     \\
\textbf{United States} 
                  & 10,614  & 51.30\%   & 46.80\%    & 41.1    & 81\%   & 8\%    & \$31,202     \\

\bottomrule
\end{tabular}
}
\end{adjustbox}
    \caption{Demographics of cities. \footnotesize Column (1) and Columns (4)-(7) show the demographics of each city according to the American Community Survey (ACS). Column (2) and Column (3) show the percentage of voters who voted for Biden (D) or Trump (R) in the 2020 Presidential Election according to data released by the Michigan Voter Information Center. (We show the 2020 Presidential Election data here since we used this data to sample cities when we conducted the study in 2023.) The last two rows present summary statistics; Column (1) and (4)-(7) are for all municipalities in the state and nation, derived from the Government Units Survey (GUS) and ACS; Column (2) and (3) are the percentage of popular votes of the state/nation. For more detailed information, see Appendix Table \ref{tab:cityFacts}.}
    \label{tab:facts_city}
\end{table}

\subsection{Operationalizing the taxonomy of local concerns and \scs{}}

\commentout{
\begin{itemize}
 
    \item \textbf{local concern:} We consider a local concern to be a type of local government operation or site of local governance and decision-making.
    For example, a local government may provide the physical and institutional infrastructure needed for providing services, maintaining public spaces, and facilitating civic engagement. Each local concern describes government operations which in a typical city may be grouped under one organization (e.g. a committee or office) although the particular organization name and structure can vary across cities.

    \item \textbf{Societal concern:} We consider a \sc{} to be a problem that influences the wellbeing, rights, or opportunities of individuals, groups, and populations. Societal concerns may be brought to the attention of a city council because \scs{} are complex and multi-faceted problems which require coordination to address.
\end{itemize}
}

Our approach is based on two dimensions, local concerns and \scs{}. To develop our taxonomies of local concerns and \scs{}, we begin with past literature \cite{mayoralSurvey, local_governments_us, Avery2016, Einstein2017}.
We then also utilized topic modeling with BERTopic \citep{bertopic} to explore the data and discover additional types of concerns and issues. We then deleted repeated topics and merged similar topics. This process resulted in  14 specific local concerns and 16 specific \scs{} (see definitions for these topics in Table \ref{tab:civic_def} and Table \ref{tab:social_def}). The first author then manually inspected a subset of 600 comments (65.72\% of annotated data, which accounts for 38.49\% of all comments) by probabilistically sampling from cities in the annotated data with weights on the number of comments of each city, each meeting, and from each speaker ID, labeling and then grouping comments in terms of local concern or \sc{}; the other authors then convened to review these taxonomies and comment labels. 
Once the taxonomies were finalized, the first author then hand-annotated the annotated data (see Figure \ref{fig:LocalGovKFold}), enabling the machine learning approach detailed below.
Whenever the author was unclear about the topic assignment for a given comment the research team would deliberate together to assign the final labels.

Examples of comments which express local concerns and \scs{} are shown in \tabref{tab:examples}.
Not all comments mention a clear local concern or \sc{} and we include this as part of our labeling scheme, for example, see C1 in \tabref{tab:examples}.
C2 is an example comment that mentions local concerns only, where the local concern is ``Zoning and rezoning and land use'' and the commenter was asking the city council to deny a conditional zoning request. C3 is an example comment that mentions \scs{} only, where the \sc{} is ``Anti-racism'', and the commenter is asking the city council to pay attention to ongoing racism issues in the city. C4 is an example comment that mentions both local concerns and \scs{}, where the local concern is ``Policing'' and the \sc{} is ``Anti-racism''. The commenter was asking the city council to approve the Driving Equity Ordinance, which is a local law in that municipality that limits the circumstances under which the police can stop drivers. 

\begin{table}[hbt!]
    \begin{adjustbox}
{width=\columnwidth,center}{
\def\sym#1{\ifmmode^{#1}\else\(^{#1}\)\fi}

\begin{threeparttable}
\begin{tabular}{p{3cm}p{15cm}}
\toprule
Neither      & C1 - On January 27, parts of the world celebrated International Holocaust Remembrance Day. Actually, that was just the official day of remembrance. The real Remembrance Day is every day of the year, and if you don't believe that, merely create for yourself a Google alert on the word Holocaust. ...                                                              \\ \addlinespace
Local Concern   & C2 - ... \textbf{The conditional zoning request should be denied} for the following reasons. It is the kind of conditional zoning that the courts have very clearly ruled against. And it would create significant possible repercussions to property owners and tenants from many blocks nearby in the area commonly known as Lower Burns Park. And there is a legal remedy. The same exact site plan could be approved in a different manner that would be legal. The way it's applied for is not, I suggest you speak with your city attorney about that for a more formal opinion... \\ \addlinespace
Societal Concern & C3 - Tonight you'll hear weeping and wailing about the \underline{Black Lives Matter} yellow paint in the park. ... But good God, is that the maximum you have to give to the black community? A dab of yellow paint on the sidewalk? \underline{That's all you're gonna give to the black community after} \underline{everything you've taken away?} This is the kind of racist city council we are dealing with.                                                                                                                                     \\ \addlinespace
Both         & C4 - I'm here to support the \textbf{Driving Equity Ordinance}. Routine traffic stops result in humiliating, traumatizing, and even worse outcomes, especially for \underline{black and brown men} who experience, as you've heard, much higher incidents. ...                                                                                                                                                                                                                                                                                                         
 \\  
\bottomrule
\end{tabular}
\begin{tablenotes}
    \item In comments with a local concern, we bold the area. In comments with a societal concern, we underline the issue. 

\end{tablenotes}
\end{threeparttable}
}
\end{adjustbox}

\commentout{Announcement      & C5 - Hello, I live here and do business here. I do hair and make and sell natural products online. So recently I decided to do some free community courses teaching small business and hair. So I did, it's gonna be at the Inkster Rec. There's various dates, and you can register online. I just wanted to let you guys know. It starts Sunday, this Sunday the 29th is on Sundays, two to four. ... & C6 -  ... I just want to remind the public that our \textbf{Friends of the Library book sale} is happening this week. It's going to be at the Radcliffe Center on Wednesday, April 19th, from 9 to 730, and on Thursday, April 20th, from 9 to 530. Thursday is going to be \$5 a bag day. You fill a bag for \$5. Every additional bag will be \$1. ...   & C7 - Hard to hear people think about the \underline{BLM} message as just words on the ground because there's so much more than that. ... I need you all to come to the Getting Real About Race workshop and lean into your microaggressions that some of you use against black people. ... & C8 - ... October is \underline{Breast Cancer Awareness Month}. ... The fire department, we're actually wearing pink T-shirts in support of breast cancer awareness. \textbf{The city police and fire department} have partnered with, together, with Friends Together. We're actually selling nice pink bracelets, \$10 a piece. All the proceeds will go to Friends Together for their cause as well.  ... So we just want to let you know that's an initiative we have collaboratively going throughout the city. So thank you.}
    \caption{Example comments spoken by public commenters at city council meetings in multiple cities of Michigan. C1 is an example comment that contains no local concern and no \sc{}. C2 is an example comment that contains a local concern (Zoning and rezoning and land use) but no \sc{}. C3 is an example comment that contains a \sc{} (Anti-racism) but no local concern. C4 is an example comment that contains both a local concern (Policing) and a \sc{} (Anti-racism). For more details, see Section \textbf{``Developing a taxonomy of local concerns and \scs{}''}. }
    \label{tab:examples}
\end{table}

Finally, we note that our focus is on comments which ask the city council to take action on a concern.
While such comments do not always state an implicit action, we consider even a request for the city council to pay attention to an issue as a kind of action. 
Since our focus is on the public's policymaking engagements with their local governments, we are not interested in announcements or comments which serve to inform the broad community, as in the following example: 

\vspace{2mm}
\begin{small}``Hello, I live here and do business here. I do hair and make and sell natural products online. So recently I decided to do some free community courses teaching small business and hair. So I did, it's gonna be at the Inkster Rec. There are various dates, and you can register online. I just wanted to let you guys know. It starts Sunday, this Sunday the 29th is on Sundays, two to four. ...''
\end{small}
\vspace{2mm}

In our data, we found that such comments occurred infrequently (12.16\% of annotated data); as such, excluding them still allowed us to characterize the vast majority of our data.

\subsection{Machine learning detection of action comment, local concerns, \& \scs{}}
In order to characterize each public comment along the local concern and \sc{} dimensions, 
we use supervised machine learning.
First, we hand-annotate comments in seven of the cities in the dataset. 
Next, we use these annotated comments 
to train machine learning models 
in order to annotate the remaining comments.

Here we propose three sets of supervised learning tasks. 
First, we predict whether a comment is addressed to the government as a call to action or a call to attention around an issue that the commenter would like to see action taken on. We term such a comment as a \commenttype{}, as it explicitly or implicitly calls the government to action. 
Next, we detect the presence of specific local concerns and \scs{}. 

\xhdr{\commenttype}
We frame the classification of \commenttype\ as a binary classification problem where 
 a comment can be either a \commenttype\  (\textsc{\commenttype}$=1$)
 or not (\textsc{\commenttype}$=0$).

 \xhdr{Local concerns}
Here we follow the taxonomy introduced above. 
For each comment, we would like to determine if it can 
be described by any of the top ten local concerns above, 
by another local concern, 
or by no local concern. 
We structure this task into a series of binary classification tasks 
and aggregate their output. First, for each comment, we predict whether it mentions any local concern. 
We assign each comment $i$  a set of local concerns $y^c_i$.
That is we assign each comment the label \textsc{Local}, 
which 
is 1 if the comment contains any local concern and 0 otherwise. 
Next, if it does have a local concern, we predict which of the ten local concerns above it contains. Let $ \mathcal{L}=\{\text{Election Administration \& Voting} ,\dots,\text{Zoning \& Rezoning}\}$ (full list in Table \ref{tab:civic_def}), 
be the set of the top ten local concerns. 
That is for each comment, for each local concern $ a \in \mathcal{L}$,
we predict whether $a$ is mentioned in comment $i$, ($C_i^a=1$). 
Thus a comment can mention multiple local concerns. 
Finally, we aggregate these predictions as follows:
\[
{
y^c_i = 
\begin{cases}
\{a \in \mathcal{L}: C_i^a=1\}, & (\textsc{Local}_i=1) \wedge (\sum_{a\in\mathcal{L}}C_i^a>0)\\
\text{Other local concern}, & (\textsc{Local}_i=1) \wedge (\sum_{a\in\mathcal{L}}C_i^a=0)\\
\text{No local concern}, & \textsc{Local}_i=0.
\end{cases}
}
\]

\begin{table}[hbt!]
    \begin{adjustbox}{width=\textwidth,center}{
\def\sym#1{\ifmmode^{#1}\else\(^{#1}\)\fi}
\begin{threeparttable}
\begin{tabular}{p{4cm}p{20cm}}
\toprule
Category                         & Definition                                                                                                                                                                                       \\ \midrule
\multicolumn{2}{l}{\textit{Top 10 \lcs}} \\ 

Election Administration and Appointments             & An explicit mention of voting, election administration, or appointments of government positions. For example, there may be explicit mentions of the term ``voting procedures'' or ``impeachment''. \\ \addlinespace

Housing                      & An explicit mention of the availability of housing, or construction around housing.  Any mention of zoning around housing would be tagged as zoning.                                                                                                                                     \\ \addlinespace
Local economy                 & An explicit mention of the local economy. For example, 
workforce education and development and business attractions and siting.                                                                                                    \\ \addlinespace
Policing                   & An explicit mention of policing. 
                                                                                          \\ \addlinespace
Public service                & An explicit mention of the provision of public services by the local government. For example, a specific service from the government, including but not limited to legal services, finance/accounting services, human resource services, social services (assistance towards particular groups), and health services. \\ \addlinespace

Public spaces and parks and recreation              & An explicit mention of public spaces. For example, parks, squares, public artworks, or public trails. Also an explicit mention of services provided by parks and recreation departments.                                                                                                                            \\ \addlinespace
Public works                 & An explicit mention of construction/installation, alteration, demolition, repair, and maintenance of local infrastructures.  This is distinct from construction around non-public housing projects for example.                                                                                                                           \\ \addlinespace
Transit corridors and parking & An explicit mention of local transit corridors and parking lots.                                                                                                                                                                                       \\ \addlinespace

Utility service             & An explicit mention of utility services, such as electricity, water, natural gas, waste management, and sewer services.                                                                                                                                 \\ \addlinespace
Zoning and rezoning          & An explicit mention of zoning and rezoning. This may or may not pertain to housing.  \\    
\midrule

\multicolumn{2}{l}{\textit{Other \lcs}}\\ 

Events and culture &  An explicit mention of city events or cultural projects such as public art, parades, and local festivals.     \\ \addlinespace 
Fire service & An explicit mention of fire services.  \\ \addlinespace
Library service &  An explicit mention of library services.  \\ \addlinespace
Transportation & An explicit mention of mobility or traffic. \\

\bottomrule
\end{tabular}

\end{threeparttable}
}
\end{adjustbox}
    \caption{ \small Definitions of local concerns. We say that a comment mentions a local concern if it references a problem, concern, or area for improvement, with an implicit or explicit reference to a city entity that would act on that issue. For example, a mention that the provision of voting stations needs to improve. }
    \label{tab:civic_def}
\end{table}

\begin{table}[htbp!]
    \begin{adjustbox}{width=\textwidth,center}{
\def\sym#1{\ifmmode^{#1}\else\(^{#1}\)\fi}
\begin{threeparttable}
\begin{tabular}{p{4cm}p{20cm}}

\toprule
Category                         & Definition                                                                                                                                                                                                                                                                                                                            \\ \midrule

\multicolumn{2}{l}{\textit{Top 10 \scs}} \\ 
Affordability                   & Concerns/awareness about the ability of individuals or households to meet their essential living expenses in the city within their income level without undue financial strain. These living expenses can be related to housing, food, healthcare, transportation, utilities, and education.                                                    \\ \addlinespace
Anti-racism                    & Concerns/awareness  about racial inequality and/or advocates for policies, practices, and attitudes that promote racial equity and justice.                                                                                                                                                                            \\ \addlinespace

Functional democracy          & Concerns about the principles of democratic governance including city charters and state constitutions, transparency, accountability, efficiency, and responsiveness of the local government.  \\ \addlinespace

Homelessness                    & Concerns/awareness around homelessness either in the city or more broadly. 
Homelessness may be brought up in the context of mental health and/or public services around this particularly vulnerable population.                                                                                 \\ \addlinespace

Incarceration and crime history & 
Concerns/awareness of mass incarceration, and discrimination against citizens with a crime history. Especially for job opportunities, rental applications, and mental health care. \\ \addlinespace

Public health                    & Concerns/awareness around public health. Public health may be brought up in the context of community health (sometimes around environmental risks),  local accessibility to physical and mental health resources, or the government’s readiness to respond to public health crises.                                                                \\ \addlinespace
Public safety                   & Concerns about criminal activities or other areas of emergencies, accidents, and hazards.                                                                                                                                                                            \\ \addlinespace
Quality of the built environment & Concerns about the quality of local buildings, public spaces, transit corridors, and other manmade infrastructure.   \\      \addlinespace

Senior, infant, child, and teenager care & Concerns/awareness of the safety, well-being, and rights of the elderly, infants, children, and teenagers.  \\ \addlinespace

Sustainability & Concerns/awareness for the health and preservation of the natural environment. May be brought up around issues of
climate change, emission and pollution, renewable energy, sustainable development, green space preservation, biodiversity, and ecosystem health.  \\ 
\midrule
\multicolumn{2}{l}{\textit{Other \scs}} \\ 
Commerce and jobs                & Concerns / awareness around the ability of the local economy to support the municipality. For example, explicit concerns about local commerce and jobs that focus on supporting small businesses, creating and increasing access to local employment opportunities, and fostering economic growth within communities by promoting entrepreneurship, investing in infrastructure, and addressing barriers such as zoning laws or access to capital. \\ \addlinespace

Culture                          &  Concerns about the culture of a municipality or community. For example, concerns that a local culture is being threatened or calls to preserve or respect the culture of a place or community.
\\ \addlinespace

Disability issues & Concerns/awareness of the safety, wellbeing, and rights of people with disabilities. \\ \addlinespace

Education &  Concerns about accessibility, quality, affordability, and inequality of education. This also includes discussions around education outcomes, budget and resource allocation, education policies, teaching technology integration, parent and community involvement, and content of education materials. \\ \addlinespace
Gender issues & Concerns/awareness around gender inequality, feminism, or LGBTQIA+ issues. \\ \addlinespace   
Natural environment and human activity & Concerns about the quality or hazards (natural disasters) of the environment, such as lakes, rivers, and forests. Also concerns about public green spaces. In contrast to the sustainability category, here we expect speakers to emphasize the effect of the natural environment on human wellbeing, rather than focus the frame on the wellbeing of the environment.\\ 

\bottomrule
\end{tabular}

\end{threeparttable}
}
\end{adjustbox}
    \caption{ \small Definitions of \scs{}. In contrast to local concerns, \scs{} do not need to connect to government entities and they should talk about problems or issues which the speaker implies. }
    \label{tab:social_def}
\end{table}

\xhdr{Societal Concerns}
Here we also follow the taxonomy introduced above.
We assign each comment $i$  a set of \scs{} $y^s_i$.
First, we for each comment we predict whether it mentions any \sc{}. 
That is we assign each comment the label \textsc{Societal}, 
which 
is 1 if the comment contains any \sc{} and 0 otherwise. 
Next, if it does have a social area we predict which of the top ten \scs{} above it contains. 
Let $\mathcal{S}=\{\text{Affordability } ,\dots,\text{Sustainability}\}$ (full list in Table \ref{tab:social_def}), 
be the set of the top ten \scs{}. 
That is for each comment, for each \sc{} $ j \in \mathcal{S}$,
we predict whether $j$ is mentioned in comment $i$, ($S_i^j=1$). 
Thus, a comment can also mention multiple \scs{}. 
Finally, we aggregate these predictions as follows:
\[
{
y^s_i = 
\begin{cases}
\{j \in \mathcal{S}: S_i^j=1\}, & (\textsc{Societal}_i=1) \wedge (\sum_{j\in\mathcal{S}}S_i^j>0)\\
\text{Other \sc{}}, & (\textsc{Societal}_i=1) \wedge (\sum_{j\in\mathcal{S}}S_i^j=0)\\
\text{No \sc{}}, & \textsc{Societal}_i=0.
\end{cases}
}
\]

\begin{figure*}[!htb]
    \centering
    \includegraphics[width=\columnwidth]{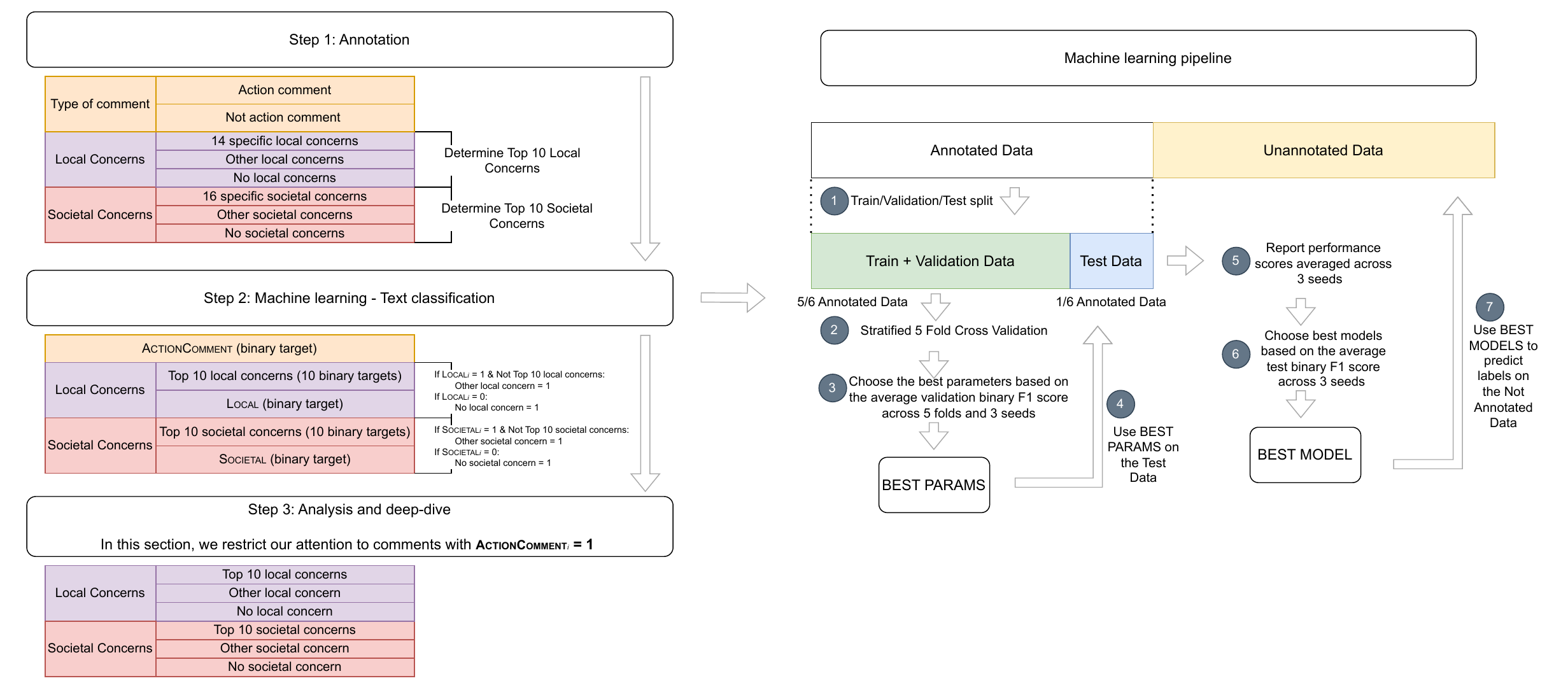}
    \caption{Project pipeline}
    \label{fig:LocalGovKFold}
\end{figure*}

\xhdr{Implementation details}
We illustrate our machine learning pipeline in Figure \ref{fig:LocalGovKFold}. Our task is to use the annotated data to predict the unannotated data. To do this, we first separate the annotated data into train, validation, and test sets, where the test set is 1/6 of the entire annotated data. For the remaining 5/6 of the annotated data, we implement a stratified 5-fold cross-validation to select the best parameters for each of the four models: Logistic Regression \& TF-IDF Vectorizer (Logistic), Support Vector Classifier \& TF-IDF Vectorizer (SVM), DistilBERT \citep{sanh2020distilbertdistilledversionbert}, and RoBERTa \citep{liu2019robertarobustlyoptimizedbert}. (See Appendix Table \ref{tab:hyperPara} for ranges used for hyperparameter tuning.) Next, we use the best parameters on the test data to find the best model for each target variable (\textsc{\commenttype}, the 10 \lcs, and the 10 \scs{}). Finally, we use the best models to predict labels on the unannotated data. This yields predicted labels for the 8 cities without annotation. 

Across all of the classification tasks of \commenttype, \textsc{Local}, and \textsc{Societal}, the input is a single public comment, which we transform into a different representation 
depending on the method. For example, when using Logistic Regression and SVMs, we represent each comment as a TI-IDF feature vector converted by TfidfVectorizer. 
When using the pre-trained language models, we input the raw text.\footnote{We ran the Logistic Regression and SVM models on 50 AMD EPYC™ 7763 64-Core processors. We ran the DistilBERT and RoBERTa models on 4 NVIDIA RTX™ A5000 graphics cards.}


\subsection{Analysis of Predictive Models}

\begin{table}[hbt!]
    \begin{adjustbox}{width=0.6\textwidth,center}{

\begin{threeparttable}
\begin{tabular}{rccc}
\toprule
  & F1     & Precision & Recall \\ \midrule
\multicolumn{1}{l}{\textsc{\commenttype}}          &        &           &        \\
\textbf{Logistic}                                                  & \textbf{0.9732} & 0.9477    & 1.0000 \\
        & (0.0000) & (0.0000)   & (0.0000) \\

\textbf{SVM}                                                       & \textbf{0.9732} & 0.9477    & 1.0000 \\
        & (0.0000) & (0.0000)   & (0.0000) \\ 

DistilBERT                                                & 0.7761 & 0.8464    & 0.7458 \\
        & (0.0074) & (0.0111)   & (0.0074) \\ 

RoBERTa                                                   & 0.8473 & 0.8640    & 0.8353 \\
        & (0.0061) & (0.0069)   & (0.0085) \\ 

\multicolumn{1}{l}{\textit{Local concerns (average across top 10)}}   &        &           &        \\
Logistic                                                  & 0.3204 & 0.4664    & 0.2994 \\
SVM                                                       & 0.1744 & 0.3172    & 0.1662 \\
DistilBERT                                                & 0.7007 & 0.7623    & 0.6851 \\
\textbf{RoBERTa}                                                   & \textbf{0.7597}  & 0.8035    & 0.7499 \\
\multicolumn{1}{l}{\textit{Societal concerns (average across top 10)}} &        &           &        \\
Logistic                                                  & 0.4908 & 0.5972    & 0.5188 \\
SVM                                                       & 0.2847 & 0.5041    & 0.2005 \\
DistilBERT                                                & 0.7870 & 0.8270    & 0.7761 \\
\textbf{RoBERTa}                                                   & \textbf{0.8130} & 0.8352    & 0.8089
      \\

\bottomrule
\end{tabular}
\begin{tablenotes}

    \item Results are averaged across 3 random seeds.
    \item Standard errors are shown in parenthesis.

\end{tablenotes}
\end{threeparttable}
}
\end{adjustbox}
    \caption{Model performance. We bold models with the highest F1 score. We use F1 score as there is class imbalance in all target variables across all the cities. For model performance on all target variables, see Appendix Table \ref{tab:results_long}. }
    \label{tab:results}
\end{table}

In \tabref{tab:results}, we see that all four models achieve a reasonable F1 score (F1 > 0.75) on the task of detecting \textsc{\commenttype}. 
For the analysis below, we use Logistic Regression. 

When it comes to detecting the presence of specific local concerns and \scs{}, the more complex RoBERTa model does best. However, we saw on one specific local concern (\textit{Public service}) and one specific \sc{} (\textit{Anti-racism}), that DistilBERT was the best model. 
Hence, we see that the tasks of detecting local concerns and \scs{} are more difficult than that of determining whether an issue is a \commenttype, and thus require models with richer representational capabilities.
For the descriptive analysis, we inspect results generated from whichever model was best for each local concern / \sc{}. For the majority of cases, this was RoBERTa.

Upon manual inspection, an incorrect categorization occurs for three major reasons: (1) topics of \lcs{} or \scs{} are related to previous commenters' discussion thus omitting context (For example, a commenter said ``I am going to talk about in support of EC4, others have talked about biking and pedestrian'', without explaining EC4 as it was explained in previous comments. For another example, a commenter said ``Now, there is a reference to 3.1.3. I'd like to let you know that I do appreciate the lines we did put down on Hubbard, but I still do want to update that for speech signs on Hubbard'' without recapping the agenda item ``3.1.3'' since this is currently being discussed); (2) low quality video recordings; and (3) commenters don't always mention specific concerns, and this can confuse the model, which speaks to existing work on citizens' framing skills in deliberative forums \citep{brady1995beyond, einstein2019local}.


\begin{table*}[hbt!]
\centering

\resizebox{\width}{!}{
    \begin{threeparttable}
\begin{adjustbox}{width=\columnwidth,center}{
\def\sym#1{\ifmmode^{#1}\else\(^{#1}\)\fi}
\begin{tabular}{lc|cccccc}
\toprule
     &                                            & \multicolumn{6}{c}{if \textsc{\commenttype},}                                                    \\ 
                  & (1) & (2)& (3)                 & (4)  &(5) &(6)                              & (7)             \\

City                  & \textsc{\commenttype} & \textsc{Local} & \textsc{Societal}                     & Top 10 \lcs    & Top 10 \scs  & Most common \lc                                             & Most common \sc                     \\ \midrule
\textit{Ground Truth} &                                                      &                                     &                                                          &                   &                   &                                                        &                                  \\ 
Ann Arbor             & 88.72\%                                              & 100.00\%                            & 83.77\%                                                  & 78.07\%           & 78.07\%           & Zoning and rezoning and land use                       & Affordability \& Anti-racism     \\
Alpena                & 89.47\%                                              & 100.00\%                            & 35.29\%                                                  & 82.35\%           & 29.41\%           & Public service                                         & Functional democracy             \\
Cedar Springs         & 90.20\%                                              & 100.00\%                            & 56.52\%                                                  & 39.13\%           & 41.30\%           & Public service                                         & Functional democracy             \\
Garden City           & 95.74\%                                              & 100.00\%                            & 71.11\%                                                  & 84.44\%           & 62.22\%           & Public service                                         & Functional democracy             \\
Inkster               & 51.35\%                                              & 100.00\%                            & 72.37\%                                                  & 88.16\%           & 60.53\%           & Public service                                         & Quality of the built environment \\
Jackson               & 87.88\%                                              & 100.00\%                            & 88.51\%                                                  & 98.28\%           & 87.36\%           & Election administration and Appointments               & Functional democracy             \\
Lansing               & 93.78\%                                              & 100.00\%                            & 95.58\%                                                  & 97.79\%           & 93.37\%           & Public service                                         & Functional democracy             \\ \midrule
\textit{Predicted}    &                                                      &                                     &                                                          &                   &                   &                                                        &                                  \\
Lathrup Village       & 85.71\%                                              & 93.75\%                             & 72.92\%                                                  & 50.00\%           & 60.42\%           & Public service                                         & Functional democracy             \\
Mount Clemens         & 72.22\%                                              & 100.00\%                            & 74.36\%                                                  & 46.15\%           & 41.03\%           & Utility service                                        & Functional democracy             \\
Pleasant Ridge        & 85.00\%                                              & 100.00\%                            & 70.59\%                                                  & 64.71\%           & 23.53\%           & Public works                                           & Sustainability                   \\
Royal Oak             & 94.92\%                                              & 97.86\%                             & 82.89\%                                                  & 65.78\%           & 60.96\%           & Transit corridors and parking                          & Functional democracy             \\
Saint Clair           & 94.87\%                                              & 100.00\%                            & 67.57\%                                                  & 56.76\%           & 78.38\%           & Transit corridors and parking                          & Quality of the built environment \\
Sterling Heights      & 97.30\%                                              & 95.83\%                             & 84.72\%                                                  & 56.94\%           & 73.61\%           & Housing                                                & Quality of the built environment \\
Saline                & 94.44\%                                              & 100.00\%                            & 96.08\%                                                  & 49.02\%           & 50.98\%           & Public works                                           & Functional democracy             \\
Williamston           & 64.00\%                                              & 100.00\%                            & 62.50\%                                                  & 25.00\%           & 43.75\%           & Public works \& Public spaces and Parks and recreation & Public safety                   
\\

\bottomrule
\end{tabular}
}
\end{adjustbox}

\begin{tablenotes}

    \item Predictive results based on majority voting across 3 random seeds.
   
\end{tablenotes}

\end{threeparttable}}

    \caption{Overview of ground truth and predicted results. \footnotesize Column (1) shows the percentage of comments with label \textsc{\commenttype}$=1$ in each city, which signals \commenttype. \textbf{Columns (2)-(7) are calculated conditional on Column (1)}. Column (2) shows the percentage of comments where at least one local concern is mentioned, given that a comment calls for an action or attention. Similarly, Column (3) shows the percentage of comments where at least one \sc{} is mentioned given the same condition. Column (4) shows the percentage of comments that discuss at least one topic in the top 10 local concerns. Column (5) shows the percentage of comments that discuss at least one topic in the top 10 \scs{}. Column (6) shows the most common local concern discussed in each city, whereas Column (7) shows the most common \sc{} discussed in each city. If there is a tie of the most common topic in Column (6) or (7), we link the topics with ``\&''. For example, there is a tie between the most common \scs{} in Ann Arbor such that there are two most common issues both ``Affordability'' and ``Anti-racism''.}
    \label{tab:manual}
    
\end{table*}

\section{A descriptive analysis of public comments}
In \tabref{tab:manual}, we summarize the distribution of the public comments in our data across local concerns and \scs{}.
As an initial result, we see that across all of the cities in our data, the majority of public comments 
can be classified as \commenttype. This shows that most of the time, people
call on the local government to take some sort of action.\footnote{The only exception is the city of Inkster; per a manual inspection, we find a larger share of comments in those meetings that announce community events rather than addressing a governance concern.}
This finding quantitatively illustrates that, beyond serving as a soapbox \citep{adams2004public}, public commenting is a venue for people to make substantive, political demands.

For the remainder of our analysis, we focus solely on comments that call for government action.
We see that across cities, the vast majority of these comments
reference a specific local concern; for most cities, \textit{Public service}, demonstrating that, at least in our dataset, individuals often express concerns regarding the quality, accessibility, or responsiveness of specific services provided by the local government. 

\begin{figure*}
\centering
  \resizebox{.95\width}{!}{
  \begin{tikzpicture}
     \input{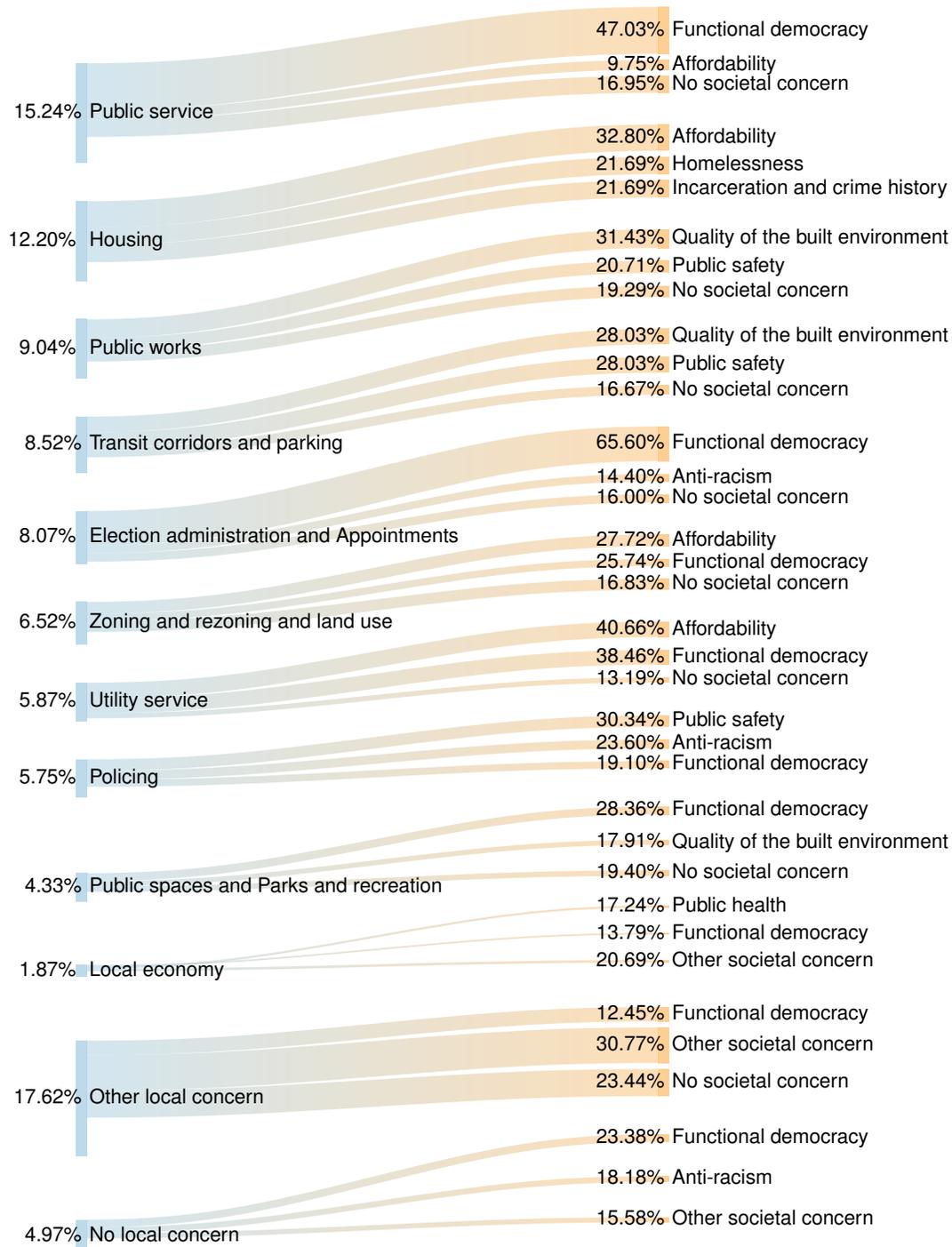}
  \end{tikzpicture}
  }
  \caption{ \footnotesize We find that overall 93.93\% of all comments include local concerns. Of these, the most common areas 
  are Public service, Housing, and Public works. We can see that the types of \scs{} which are brought up in each local concern vary. For example, functional democracy is commonly mentioned both in public service and election administration, but in housing we see that affordability is the most common theme.}
  \label{fig:sankey_civic}
\end{figure*}

The prevalence of \scs{} is lower, with greater variation across cities; in particular, comments in some of the larger cities (Ann Arbor, Jackson, Lansing), 
mention \scs{} more often than smaller cities (Alpena, Cedar Springs, Willamston). 
The predominant \sc{} across cities is \textit{Functional democracy}, illustrating that---at least in our data---people frequently raise procedural concerns about governance.

\begin{figure*}
  \centering
    \resizebox{.95\width}{!}{ 
  \begin{tikzpicture}
  \input{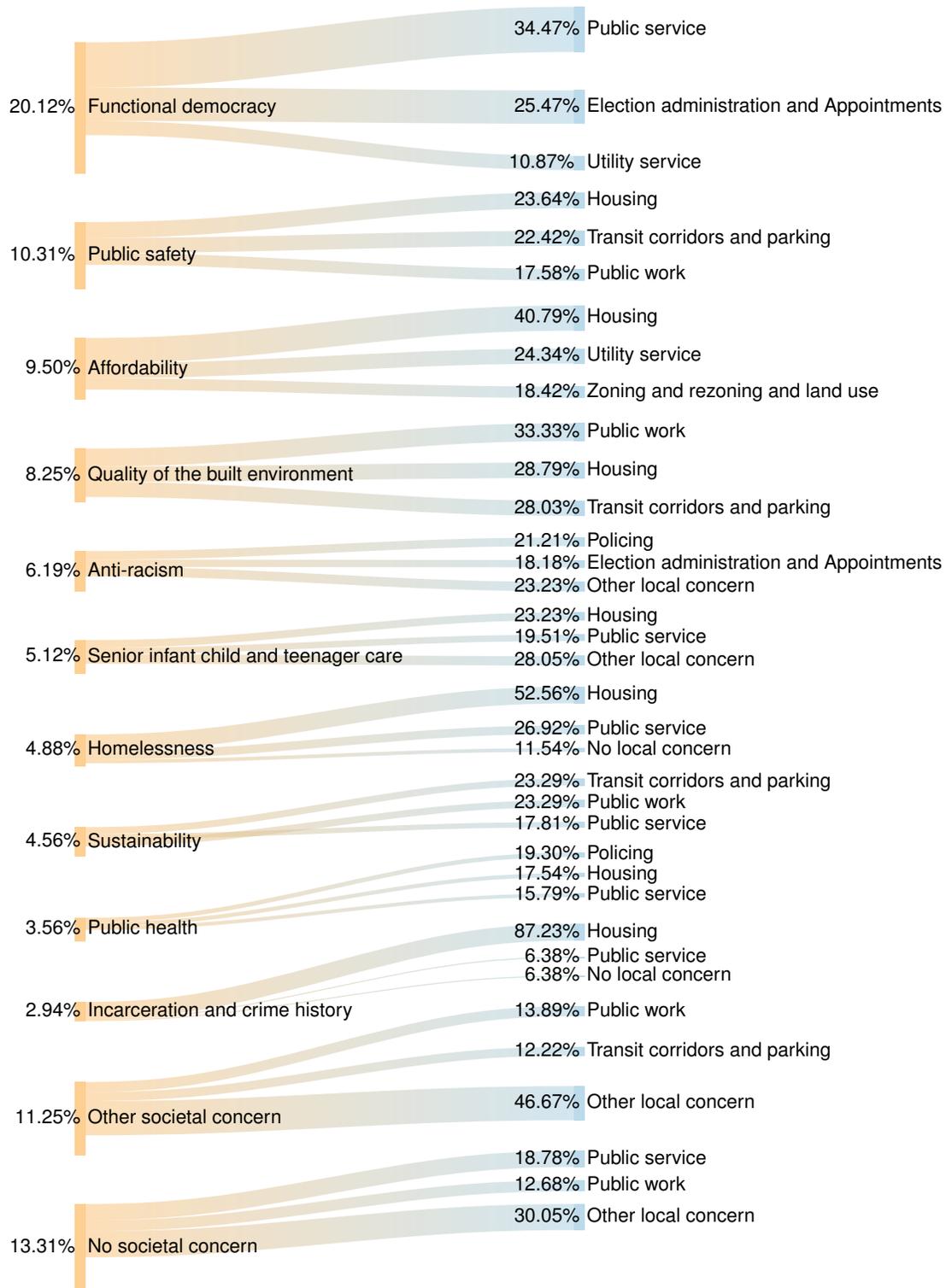}
  \end{tikzpicture}
  }
  \caption{\footnotesize We find that overall 83.22\% of all comments include \scs{}. Of these, the most common areas 
  are Functional Democracy, Public Safety, and Affordability. The local concern of housing occurs across different \scs{}, while others (such as \textit{Public service} and Election administration) are more tightly associated with a particular issue. }
  \label{fig:sankey_social}
\end{figure*}

We visualize the frequency of the most common local concerns and \scs{}, as well as how these dimensions interact, in 
\figref{fig:sankey_civic} and \figref{fig:sankey_social}.
\figref{fig:sankey_civic} begins with local concerns, and for each local concern displays the top three \scs{}. \figref{fig:sankey_social} begins with \scs{}, and for each \sc{} displays the most common local concerns. 
In each figure, the width of the edges represents the number of comments. Percentages are calculated conditional on the respective node in the upper level. 

In \figref{fig:sankey_civic}, we see that commenters can draw on multiple \scs{} when articulating a policy demand in a particular local concern.
For example, 
when talking about \textit{Public works} (9.04\% of all \commenttype), commenters may cite concerns about the \textit{Quality of the built environment} or \textit{Public safety},
perhaps reflecting slightly diverging policy concerns (e.g., calls for more green spaces downtown vs complaints about a run-down public pool) or values.
Commenters who mention the local concern of \textit{Policing} (5.75\% of \commenttype)
often raise the \scs{} of \textit{Public safety} or \textit{Anti-racism}, 
reflecting broader discourses that cast police forces as necessary for safety or as fundamentally racist (e.g., \citep{vitale2021end}).

\figref{fig:sankey_social} shows the distribution of local concerns referenced for comments mentioning each \sc{}.
Here, we see that commenters can map a broad concern onto multiple concrete areas of local governance and policymaking.
For instance, concerns around the \textit{Quality of the built environment} (8.25\% of \commenttype) are often situated in public works projects as well as in housing or transit,
while concerns around \textit{Functional democracy} are more concentrated in matters of \textit{Public service}.

We now provide examples of how our framework can be used to examine the \lcs{} and \scs{} that are mentioned in public comments,
as well as the relationships between these two dimensions.
We choose the \lc{} of \textit{Housing}, which is an
extensively studied concern \citep{Desmond2017,Hankinson2018,Holleran2022,Schwartz2014,anzia2022local}, 
and the \sc{} of \textit{Anti-racism} which is a broadly studied \sc{} in local politics \citep{williams2021blacklivesmatter,hoang2024defund,sances2023defund}. We show two example comments from the \lc{} of \textit{Housing} and the \sc{} of \textit{Anti-racism} in Table \ref{tab:deep_dive_examples}.

\begin{table}[hbt!]
    \begin{adjustbox}{width=.8\columnwidth,center}{
\def\sym#1{\ifmmode^{#1}\else\(^{#1}\)\fi}

\begin{tabular}{lccccc}
\toprule
      & Comments &Top 10 & Top 10  &  & Anti-\\ 
City                                                       &  count & local concerns & \scs &  Housing & racism \\   
    \midrule

\textit{Ground Truth} \\
Ann Arbor        & 257                                & 178                                    & 196                                      & 29                          & 55                              \\
Alpena           & 19                                 & 14                                     & 6                                        & 1                           & 0                               \\
Cedar Springs    & 51                                 & 18                                     & 20                                       & 0                           & 0                               \\
Garden City      & 47                                 & 38                                     & 29                                       & 2                           & 0                               \\
Inkster          & 148                                & 67                                     & 77                                       & 3                           & 1                               \\
Jackson          & 198                                & 171                                    & 171                                      & 40                          & 29                              \\
Lansing          & 193                                & 177                                    & 175                                      & 66                          & 11                              \\ \midrule
\textit{Predicted} \\
Lathrup Village  & 56                                 & 26                                     & 32                                       & 0                           & 0                               \\
Mount Clemens    & 54                                 & 21                                     & 20                                       & 3                           & 0                               \\
Pleasant Ridge   & 20                                 & 11                                     & 5                                        & 1                           & 1                               \\
Royal Oak        & 197                                & 123                                    & 116                                      & 28                          & 0                               \\
Saint Clair      & 39                                 & 21                                     & 30                                       & 2                           & 0                               \\
Sterling Heights & 74                                 & 41                                     & 53                                       & 18                          & 0                               \\
Saline           & 54                                 & 27                                     & 27                                       & 0                           & 2                               \\
Williamston      & 25                                 & 4                                      & 9                                        & 0                           & 0                              
   \\ \bottomrule                         
\end{tabular}
}
\end{adjustbox}
    \caption{Annotation and prediction by city, numbers refer to comments that contain this local concern or \sc{}. For details on occurrences of all local concerns and \scs{}, see Appendix Table \ref{tab:anno_city}.}
    \label{tab:deep_dive}
\end{table}

\subsection{Housing}
The provision and development of housing is a key area of concern for local governments \citep{Desmond2017,Hankinson2018,Holleran2022,Schwartz2014,anzia2022local}.
Indeed, much of the past scholarship on local public commenting has focused on comments solicited in the course of policymaking decisions around housing development in cities like San 
 Francisco and Boston \citep{einstein2019local,yoder2020a,sahn2024public}.
Our framework can build on this research by quantifying the prevalence of housing-related comments relative to other local concerns, and by comparing across multiple cities.

Per \figref{fig:sankey_civic}, we see that housing is the second most common local concern in our data, underlining its importance in local policymaking.
However, as shown in Table \ref{tab:deep_dive}, in some cities (e.g., Inkster and Lathrup Village) housing is barely mentioned, if at all.
This contrast potentially points to interesting differences in commenters' concerns across cities which may or may not receive large influxes of new residents, which future work could fruitfully take up.

Per \figref{fig:sankey_civic}, housing comments tend to mention \textit{affordability}, \textit{homelessness} and \textit{incarceration and crime history} as \scs{}. 
This could reflect that the category spans multiple related policy areas (e.g., the availability of affordable housing vs background checks for renters). 
Such statistics could also point to the diverging frames articulated by different commenters (e.g., universal access to housing vs safety and control of public space). 
As such, our framework potentially points to sites of conflict that future work could study in more depth.

\begin{table}[hbt!]
    \begin{adjustbox}{width=\columnwidth,center}{
\def\sym#1{\ifmmode^{#1}\else\(^{#1}\)\fi}

\begin{threeparttable}
\begin{tabular}{p{3cm}p{15cm}}
   \toprule

Housing                          &     ...The right of every Lansing citizen  is a safe and affordable place to live. Not every citizen has this and especially returning citizens. Those who have been previously incarcerated are labeled and stigmatized as not worthy of a fair chance at housing in Lansing. ...  We know that safe and affordable housing gives a returning citizen a chance to make a new life and to be a contributing member of the community. I'm asking this city council to take action to pass a fair chance housing ordinance that last year's council did not do. ...            \\ \addlinespace
  &   ...the Renter's Commission has advocated for sustainability, as has the Environmental Commission and the Energy Commission, and they all have resolutions before you dealing with natural gas for both heating and for appliances. And I'm sure you've read the details of those various resolutions. Whatever we do, we must do something more than what we are currently doing. And we must do so promptly. As they ask of you, please treat this  with the great urgency fitting a declared emergency. And as they ask, please, with equal urgency, implement carrot and stick measures to promote sustainable construction.  ...              \\ \midrule
Anti-racism                      &  ... There is extensive evidence that police and policing cause harm to black, immigrant, indigenous, queer people and other people of color, as does involvement in the criminal legal system. ...    \\ \addlinespace
 &  ... To allow an employee of the city to post inflammatory, derogatory,  and racist comments on a public platform and to go unanswered in a city literally founded on the abolition of slavery is wrong. If city council is silent on this topic, in my opinion it stands as condoning her published statements and questions the council's moral authority to lead this city with a proud history founded on freedom for all people. ... \\ \bottomrule
\end{tabular}

\end{threeparttable}

}
\end{adjustbox}
    \caption{We show two example comments from the local concern of \textit{Housing} and the \sc{}{} of \textit{Anti-racism}. 
    The first comment in \textit{Housing} discusses how government should take action so that citizens with \textit{Incarceration and crime history} can have a fair chance in applying for \textit{affordable} housing. The second comment in \textit{Housing} urges the government to promote \textit{Sustainability} in housing planning and construction process. The first comment in \textit{Anti-racism} raises public awareness to racism in \textit{Policing}. The second comment in \textit{Anti-racism} asks the government to regulate \textit{Public service} in order to ensure \textit{Functional democracy}.
    }
    \label{tab:deep_dive_examples}
\end{table}

\subsection{Anti-racism}
Research has traced how the Black Lives Matter movement in recent years has foregrounded the issue of anti-racism, especially in relation to local police departments \citep{williams2021blacklivesmatter,hoang2024defund,sances2023defund}. Indeed, per \figref{fig:sankey_social} we see that the most frequent local concern associated with comments in our data that bring up \textit{Anti-racism} is \textit{Policing} (21.21\%). 
However, we find that other local concerns are mentioned as well, notably \textit{Election administration and Appointments} (18.18\%), perhaps pointing to the composition of governing bodies as another site considered by commenters for rectifying racial injustices (see Table \ref{tab:deep_dive_examples} for an example).
Future work could elaborate on the ways that people articulate anti-racist concerns in terms of local-level policies;
a particularly interesting line of research could examine longitudinal shifts as different social movements (e.g., the Black Lives Matter movement, tenants' unions, environmental justice groups) seek to locate race as a central issue across local concerns.

While anti-racism is only the fifth most common \sc{}{} mentioned, occurring in 6.19\% of all comments, we see that its occurrence varies across cities. 
Per Table \ref{tab:deep_dive}, 21\% of comments in Ann Arbor mention \textit{anti-racism}, making it among the most frequently-cited \scs{}{} in that city. 
Future work could build on this observation to characterize and explain variation across places; for instance, we may speculate that the prevalence of \textit{anti-racism} reflects Ann Arbor's relatively young population.

\section{Discussion}
Relying on YouTube videos, we provide a computational framework that can comprehensively examine public commenting activity across the diversity of concerns that people bring up during city council meetings. First, our results reinforce the benefit of using YouTube for retaining public meeting data. The relative consistency across YouTube videos, for example, \citep[e.g.][]{bararisimko2024} allows us to better extract comments for comparison across multiple municipalities.  In turn, this type of comparison can better scale to track national and over-time variation. 

Next, our framework proposes taxonomies on two dimensions of public comments---local concerns and \scs{}---alongside a computational pipeline to derive and infer these taxonomies across a large dataset of public comments. This framework bridges research on local politics with methods previously applied to the study of social media and digital trace data. We illustrate how our framework can be applied to a dataset of public comments across cities in Michigan.

Through our two dimensions, we discover a rich and varied landscape of public concerns.  
The comments in our data brought up a range of local concerns,
 reflecting matters like \textit{Public service}, \textit{Housing}, and
\textit{Public works} that local governments are typically charged with \citep{anzia2022local}. 
The comments also brought up a range of \scs{}, spanning such concerns as \textit{Functional democracy}, \textit{Public safety}, and \textit{Anti-racism}.
Our framework also surfaced ways that local concerns and \scs{} interact. 

This work contributes to the study of participation in local politics in several ways. 
First, we show that public comments are not always centered on local concerns, 
a finding with relevance to how we think about local engagement. 
In some cities (Inkster), commenters were more likely 
to discuss \scs{} than particularly local concerns. 
For example, commenters would take the time to speak about their general concerns with social injustice and racism, sometimes without directly connecting these issues
to local affairs.

Second, we see very little evidence that public comment periods are flooded by 
``soap boxers'' -- people speaking at length without any connection or relevance to local governance \citep{adams2004public}. 
While there were a few cases of commenters engaging in this type of behavior, such comments occurred very infrequently. 
The majority of the time, participants commented on concrete local concerns
or on \scs{} which they wanted more government action and accountability 
around. Nearly 90\% of the time, comments called their government officials 
to some type of action. For example, ``I ask that the PUD not be granted'' is calling the government to deny a proposal on the agenda items, and ``Dissolve the Council of the Commons so they stop wasting staff time and the time of the two council members sitting on that body'' is asking the government to discontinue a council that supervises the planning of a public space. 

Broadly, our results also hint at the possibility of nationalization in local government \citep{anzia2021, hopkins2022parties}. Although we do not track nationalization directly, we do see mentions of broad issues that ask local governments to engage in governance beyond their jurisdictions. Broadly speaking, it is possible that participants use the local space -- which could, in theory, offer more direct access to politicians --  to confront issues that seem hopeless from the perspective of national (partisan)
politics. 
For example, an especially striking issue emerging in our data was the discussions around the Palestinian-Israeli conflict. 
In theory, this would not be considered the type of concern brought to local government, as international conflicts are squarely in the jurisdiction of the federal government. 
However, we saw that people participating in public meetings were not only able to effectively use the public comment space to issue direct concerns to their local officials, they also transformed the broad \sc{} into a concrete local action:
that of a demand for a ceasefire resolution. For instance, 45 out of 51 comments were about ceasefire at the Ann Arbor city council meeting on Dec 18, 2023. Similarly, 12 out of 21 comments were about ceasefire at the Lansing city council meeting on Dec 11, 2023.
While a ceasefire resolution from the city council of a relatively 
small city, may seem like a purely politically symbolic act, it also serves as an example of the ways in which local political spaces can afford the translation of concern into demand and the ways in which the public 
can utilize the spaces which are made available to them. We hope that our framework lays the groundwork for investigating these spaces, and for proposing and answering new questions around participation in local politics.

\subsection{Limitations and future work}
Our present work is constrained to our relatively small dataset of 15 cities in the state of Michigan.
In ongoing work, we are expanding our efforts to collect data of council meetings across the United States.
We note that while our framework can be applied to any dataset,
a future approach could also explore unsupervised methods for inferring taxonomies\footnote{We did explore such an approach and found that it lacked reliable quality and structure. However, these approaches may be more appealing as technology evolves.}.
In our ongoing work, we also aim to determine the extent to which the machine learning models we've trained generalize across more places.

While we illustrate how our framework enables analyses at the level of content categories, future work could consider more fine-grained analyses of variations in language use within and across different local concerns and \scs{},
or discourse analysis approaches \citep{tracy2007speaking}.
Future work could also clarify the relationship between these two dimensions, building on the joint statistics our framework can compute.
For instance, do different \scs{} associated with the same local concern reflect differing framing strategies, or do they point to differing---if thematically related---policies? 
Conversely, do different local concerns associated with the same \sc{} reflect differing policymaking strategies for enacting a broader concern such as anti-racism or sustainability? 

Our work does not consider the identity of commenters, and hence questions regarding their representativeness. 
Future work could enrich existing studies of representativeness \citep{einstein2019local,sahn2024public} using our framework:
for instance, while past work has noted demographic differences in commenters who are pro- versus anti-housing development,
other empirical studies could measure such differences for other concerns, as well as consider the propensity of different types of commenters (e.g., old versus young, homeowner versus renter) to bring up particular local concerns or \scs{}.


Finally, as a descriptive approach, our framework leaves open the question of whether mentioning certain local concerns or \scs{} leads to policy outcomes. 
Nonetheless, we suggest that our approach could scaffold future studies of commenting effectiveness: for instance, for a given local concern, are certain \scs{} more compelling to policymakers than others?


\bibliography{main}

\section{Appendix}

\subsection{City Facts}

In \tabref{tab:cityFacts} we see descriptive data for the cities included in our dataset, where we use the US Census American Community Survey and the Michigan Voter Information Center to compute statistics. We can see the cities range in size from 2610 (Pleasant Ridge) to 134062 (Sterling Heights) in population. While the average political leaning across the cities skews Democrat, this skew is much less severe than if we were to consider the largest cities, or all cities with available recordings of city council meetings. This selection of cities brings the average political leaning as close as possible to the leaning of the state of Michigan, given the cities which publish meetings.

\begin{table*}[hbt!]
    \begin{adjustbox}{width=\columnwidth,center}{
\def\sym#1{\ifmmode^{#1}\else\(^{#1}\)\fi}

\begin{tabular}{lccccccccccccccc}
\toprule
\textbf{}                                    & Ann Arbor    & Alpena       & Cedar Springs & Garden City  & Inkster     & Jackson     & Lansing      & Lathrup Village & Mt Clemens   & Pleasant Ridge & Royal Oak    & St Clair     & Sterling Heights & Saline       & Williamston  \\ \midrule
square miles                                 & 28.2         & 8.2          & 2             & 5.9          & 6.3         & 10.8        & 39.1         & 1.5             & 4.1          & 0.6            & 11.8         & 2.9          & 36.4             & 4.3          & 2.5          \\
population (2021)                            & 122731       & 10181        & 3635          & 27268        & 25849       & 31810       & 113592       & 4098            & 15735        & 2610           & 58368        & 5489         & 134062           & 9072         & 3845         \\
population (2022)                            & 122216       & 10200        & 3646          & 27203        & 25839       & 31420       & 112986       & 4076            & 15676        & 2609           & 58053        & 5496         & 133744           & 8973         & 3810         \\
county                                       & Washtenaw    & Alpena       & Kent          & Wayne        & Wayne       & Jackson     & Ingham       & Oakland         & Macomb       & Oakland        & Oakland      & St Clair     & Macomb           & Washtenaw    & Ingham       \\
2020 \# votes for Biden                      & 59210        & 2339         & 558           & 6923         & 9415        & 7015        & 37439        & 2702            & 4465         & 1065           & 25837        & 1304         & 30591            & 3820         & 1,261        \\
2020 \# votes for Trump                      & 7662         & 2936         & 956           & 7929         & 941         & 5043        & 12536        & 398             & 3157         & 381            & 12890        & 2057         & 38452            & 2096         & 1,033        \\
2020 \# presidential votes                   & 67559        & 5393         & 1553          & 15118        & 10467       & 12332       & 51361        & 3123            & 7767         & 2017           & 39389        & 3427         & 69797            & 6020         & 2,326        \\
2020 \# votes for independent candidates     & 687          & 118          & 39            & 266          & 111         & 274         & 1386         & 23              & 145          & 571            & 662          & 66           & 754              & 104          & 32           \\
\% democrat votes                            & 87.64\%      & 43.37\%      & 35.93\%       & 45.79\%      & 89.95\%     & 56.88\%     & 72.89\%      & 86.52\%         & 57.49\%      & 52.80\%        & 65.59\%      & 38.05\%      & 43.83\%          & 63.46\%      & 54.21\%      \\
\% republican votes                          & 11.34\%      & 54.44\%      & 61.56\%       & 52.45\%      & 8.99\%      & 40.89\%     & 24.41\%      & 12.74\%         & 40.65\%      & 18.89\%        & 32.72\%      & 60.02\%      & 55.09\%          & 34.82\%      & 44.41\%      \\
\% independent vote                          & 1.02\%       & 2.19\%       & 2.51\%        & 1.76\%       & 1.06\%      & 2.22\%      & 2.70\%       & 0.74\%          & 1.87\%       & 28.31\%        & 1.68\%       & 1.93\%       & 1.08\%           & 1.73\%       & 1.38\%       \\
2022 state general election votes            & 55639        & 4441         & 1180          & 11738        & 6797        & 8756        & 39472        & 2724            & 5867         & 1881           & 33781        & 2862         & 52967            & 5339         & 2,064        \\
2022 state election turnout                  & 53.18\%      & 53.24\%      & 50.32\%       & 54.93\%      & 32.78\%     & 40.10\%     & 48.91\%      & 70.39\%         & 51.22\%      & 77.44\%        & 68.17\%      & 65.88\%      & 59.50\%          & 72.95\%      & 69.78\%      \\
2020 state general election votes            & 67989        & 5427         & 1562          & 15195        & 10529       & 12411       & 52103        & 3143            & 7810         & 2033           & 39658        & 3452         & 70209            & 6067         & 2351         \\
2020 state election turnout                  & 64.98\%      & 65.06\%      & 66.61\%       & 71.11\%      & 50.78\%     & 56.83\%     & 64.56\%      & 81.21\%         & 68.18\%      & 83.70\%        & 80.03\%      & 79.47\%      & 78.87\%          & 82.89\%      & 79.48\%      \\
\# registered voters                         & 104633       & 8341         & 2345          & 21369        & 20733       & 21837       & 80706        & 3870            & 11455        & 2429           & 49554        & 4344         & 89013            & 7319         & 2958         \\
\# active voters                             & 83727        & 6952         & 2316          & 18284        & 16674       & 18895       & 68187        & 3471            & 10543        & 2149           & 43427        & 3936         & 78085            & 6628         & 2681         \\
\% active voters                             & 80.02\%      & 83.35\%      & 98.76\%       & 85.56\%      & 80.42\%     & 86.53\%     & 84.49\%      & 89.69\%         & 92.04\%      & 88.47\%        & 87.64\%      & 90.61\%      & 87.72\%          & 90.56\%      & 90.64\%      \\
median age                                   & 25.9         & 44.7         & 41.8          & 40.9         & 32.3        & 35          & 33.9         & 48.2            & 40.2         & 43.9           & 35.9         & 41.9         & 40.7             & 44.6         & 40.2         \\
\%white                                      & 68.00\%      & 93.00\%      & 88.00\%       & 85.00\%      & 16.00\%     & 65.00\%     & 52.00\%      & 30.00\%         & 65.00\%      & 91.00\%        & 83.00\%      & 95.00\%      & 80.00\%          & 86.00\%      & 97.00\%      \\
\%black                                      & 6.00\%       & 2.00\%       & 0.00\%        & 6.00\%       & 75.00\%     & 20.00\%     & 22.00\%      & 64.00\%         & 24.00\%      & 1.00\%         & 5.00\%       & 2.00\%       & 6.00\%           & 2.00\%       & 0.00\%       \\
\%asian                                      & 15.00\%      & 0.00\%       & 0.00\%        & 1.00\%       & 1.00\%      & 1.00\%      & 6.00\%       & 0.00\%          & 0.00\%       & 1.00\%         & 4.00\%       & 0.00\%       & 9.00\%           & 4.00\%       & 0.00\%       \\
\%multi-racial                               & 5.00\%       & 2.00\%       & 3.00\%        & 3.00\%       & 6.00\%      & 7.00\%      & 8.00\%       & 2.00\%          & 7.00\%       & 3.00\%         & 3.00\%       & 1.00\%       & 3.00\%           & 5.00\%       & 2.00\%       \\
\%hispanic                                   & 5.00\%       & 4.00\%       & 8.00\%        & 5.00\%       & 2.00\%      & 6.00\%      & 11.00\%      & 4.00\%          & 3.00\%       & 4.00\%         & 4.00\%       & 2.00\%       & 2.00\%           & 3.00\%       & 1.00\%       \\
per\_capita\_income                          & \$50,788.00  & \$30,936.00  & \$25,042.00   & \$31,043.00  & \$22,420.00 & \$23,447.00 & \$29,023.00  & \$54,461.00     & \$31,939.00  & \$84,370.00    & \$59,760.00  & \$44,068.00  & \$35,742.00      & \$44,741.00  & \$51,635.00  \\
median\_household\_income                    & \$78,740.00  & \$43,613.00  & \$47,300.00   & \$63,630.00  & \$38,381.00 & \$41,988.00 & \$48,962.00  & \$97,750.00     & \$55,154.00  & \$164,861.00   & \$92,799.00  & \$71,771.00  & \$78,049.00      & \$88,388.00  & \$76,086.00  \\
\%under \$50k                                & 35.00\%      & 56.00\%      & 52.00\%       & 39.00\%      & 59.00\%     & 58.00\%     & 51.00\%      & 12.00\%         & 47.00\%      & 12.00\%        & 26.00\%      & 30.00\%      & 33.00\%          & 25.00\%      & 33.00\%      \\
\% \$50k - \$100k                              & 25.00\%      & 30.00\%      & 34.00\%       & 35.00\%      & 31.00\%     & 28.00\%     & 33.00\%      & 39.00\%         & 32.00\%      & 16.00\%        & 28.00\%      & 32.00\%      & 28.00\%          & 33.00\%      & 31.00\%      \\
\% \$100k - \$200k                             & 25.00\%      & 11.00\%      & 13.00\%       & 24.00\%      & 8.00\%      & 11.00\%     & 15.00\%      & 37.00\%         & 19.00\%      & 36.00\%        & 32.00\%      & 27.00\%      & 32.00\%          & 33.00\%      & 20.00\%      \\
\%over \$200k                                & 16.00\%      & 2.00\%       & 1.00\%        & 2.00\%       & 3.00\%      & 2.00\%      & 2.00\%       & 13.00\%         & 2.00\%       & 36.00\%        & 14.00\%      & 12.00\%      & 7.00\%           & 9.00\%       & 15.00\%      \\
\%below\_poverty                             & 23.10\%      & 20.90\%      & 24.70\%       & 10.80\%      & 35.10\%     & 24.00\%     & 22.20\%      & 6.10\%          & 16.80\%      & 2.60\%         & 6.50\%       & 9.10\%       & 9.70\%           & 5.10\%       & 7.30\%       \\
\#households                                 & 49337        & 4903         & 1537          & 11025        & 9222        & 13131       & 53147        & 1574            & 6583         & 1149           & 28986        & 2201         & 49804            & 3737         & 1753         \\
\#persons per household                      & 2.2          & 2            & 2.3           & 2.5          & 2.8         & 2.3         & 2.1          & 2.6             & 2.2          & 2.3            & 2            & 2.5          & 2.7              & 2.4          & 2.2          \\
\#housing\_units                             & 53113        & 5446         & 1617          & 11324        & 11081       & 14797       & 55954        & 1632            & 6888         & 1190           & 31204        & 2306         & 51797            & 3954         & 1862         \\
occupancy \%                                 & 93.00\%      & 90.00\%      & 95.00\%       & 97.00\%      & 83.00\%     & 89.00\%     & 95.00\%      & 96.00\%         & 96.00\%      & 97.00\%        & 93.00\%      & 95.00\%      & 96.00\%          & 95.00\%      & 94.00\%      \\
\%occupancy by owner                         & 43.00\%      & 64.00\%      & 57.00\%       & 81.00\%      & 44.00\%     & 55.00\%     & 56.00\%      & 91.00\%         & 60.00\%      & 95.00\%        & 65.00\%      & 82.00\%      & 76.00\%          & 71.00\%      & 69.00\%      \\
\%single-unit                                & 50.00\%      & 70.00\%      & 57.00\%       & 91.00\%      & 76.00\%     & 69.00\%     & 65.00\%      & 97.00\%         & 70.00\%      & 98.00\%        & 69.00\%      & 86.00\%      & 76.00\%          & 74.00\%      & 64.00\%      \\
\%multi-unit                                 & 50.00\%      & 30.00\%      & 29.00\%       & 9.00\%       & 23.00\%     & 31.00\%     & 32.00\%      & 2.00\%          & 28.00\%      & 2.00\%         & 31.00\%      & 18.00\%      & 22.00\%          & 25.00\%      & 24.00\%      \\
median value of owner-occupied housing units & \$435,000.00 & \$105,400.00 & \$188,600.00  & \$152,800.00 & \$69,000.00 & \$88,000.00 & \$122,400.00 & \$261,800.00    & \$150,900.00 & \$393,900.00   & \$289,800.00 & \$198,700.00 & \$272,900.00     & \$302,200.00 & \$199,600.00 \\
\%no degree                                  & 3.00\%       & 8.00\%       & 9.00\%        & 9.00\%       & 16.00\%     & 13.00\%     & 11.00\%      & 5.00\%          & 13.00\%      & 1.00\%         & 4.00\%       & 5.00\%       & 11.00\%          & 3.00\%       & 2.00\%       \\
\%high school                                & 5.00\%       & 35.00\%      & 42.00\%       & 37.00\%      & 37.00\%     & 38.00\%     & 25.00\%      & 19.00\%         & 35.00\%      & 5.00\%         & 13.00\%      & 32.00\%      & 26.00\%          & 13.00\%      & 14.00\%      \\
\%some college                               & 13.00\%      & 43.00\%      & 39.00\%       & 40.00\%      & 35.00\%     & 34.00\%     & 35.00\%      & 27.00\%         & 36.00\%      & 16.00\%        & 22.00\%      & 37.00\%      & 29.00\%          & 27.00\%      & 35.00\%      \\
\%bachelors                                  & 33.00\%      & 8.00\%       & 8.00\%        & 10.00\%      & 9.00\%      & 11.00\%     & 17.00\%      & 24.00\%         & 12.00\%      & 40.00\%        & 37.00\%      & 18.00\%      & 21.00\%          & 27.00\%      & 28.00\%      \\
\%post-grad                                  & 46.00\%      & 6.00\%       & 4.00\%        & 4.00\%       & 3.00\%      & 5.00\%      & 12.00\%      & 26.00\%         & 5.00\%       & 37.00\%        & 24.00\%      & 8.00\%       & 12.00\%          & 31.00\%      & 21.00\%     \\ \bottomrule
\end{tabular}

}
\end{adjustbox}
    \caption{City Facts}
    \label{tab:cityFacts}
\end{table*}

\begin{figure*}[!htb]
    \includegraphics[width=\columnwidth]{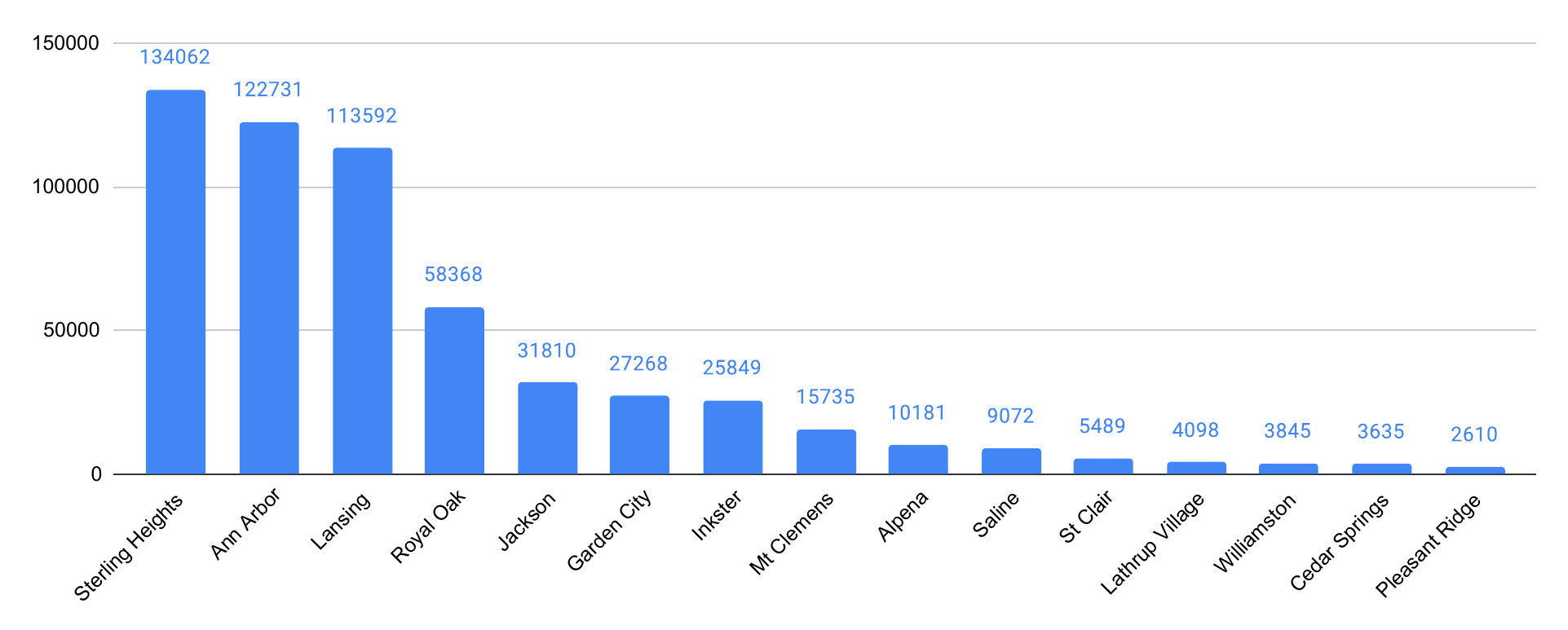}
    \caption{Population distribution of selected Michigan cities in 2021. Ranked from largest to smallest population based on data from the US Census American Community Survey (ACS).}
    \label{fig:population}
\end{figure*}

\begin{table*}[hbt!]
    \begin{adjustbox}{width=\columnwidth,center}{
\def\sym#1{\ifmmode^{#1}\else\(^{#1}\)\fi}
\begin{tabular}{l|ccccccc|cccccccc} 
\toprule

    &  \multicolumn{7}{c}{Ground Truth}                                                           & \multicolumn{8}{c}{Predicted}   \\ 
   & Ann  &  & Cedar  & Garden  &  &  &  & Lathrup  & Mount  & Pleasant  & Royal  & Saint  & Sterling  &  &  \\
City                                     &  Arbor & Alpena &  Springs &  City & Inkster & Jackson & Lansing &  Village &  Clemens &  Ridge &  Oak &  Clair &  Heights & Saline & Williamston \\

\midrule
Comments count         & 257 & 19 & 51 & 47 & 148 & 198 & 193 & 56 & 54 & 20 & 197 & 39 & 74 & 54 & 25 \\
\textsc{\commenttype}       & 228 & 17 & 46 & 45 & 76  & 174 & 181 & 48 & 39 & 17 & 187 & 37 & 72 & 51 & 16 \\

\textsc{Civic}         & 228 & 17 & 46 & 45 & 76  & 174 & 181 & 49 & 49 & 18 & 190 & 38 & 70 & 52 & 17 \\
\textsc{Social}         & 212 & 7  & 27 & 33 & 89  & 173 & 179 & 36 & 33 & 12 & 160 & 26 & 61 & 50 & 11 \\

Top 10 civic areas          & 178 & 14 & 18 & 38 & 67  & 171 & 177 & 26 & 21 & 11 & 123 & 21 & 41 & 27 & 4  \\
Top 10 social issues        & 196 & 6  & 20 & 29 & 77  & 171 & 175 & 32 & 20 & 5  & 116 & 30 & 53 & 27 & 9  \\

\midrule
\textit{Civic Areas}                     &           &        &               &             &         &         &         &                 &               &                &           &             &                  &        &             \\
Housing                                  & 29        & 1      & 0             & 2           & 3       & 40      & 66      & 0               & 3             & 1              & 28        & 2           & 18               & 0      & 0           \\
Public service                           & 33        & 10     & 13            & 17          & 21      & 29      & 70      & 17              & 4             & 2              & 19        & 1           & 4                & 8      & 2           \\
Election administration and Appointments & 6         & 0      & 2             & 1           & 6       & 60      & 28      & 9               & 4             & 0              & 3         & 0           & 7                & 4      & 1           \\
Utility service                          & 14        & 0      & 1             & 8           & 12      & 23      & 6       & 3               & 5             & 1              & 3         & 4           & 5                & 6      & 0           \\
Policing                                 & 26        & 1      & 3             & 2           & 9       & 17      & 19      & 0               & 1             & 0              & 5         & 1           & 4                & 3      & 2           \\
Public works                             & 13        & 0      & 1             & 8           & 14      & 16      & 11      & 9               & 2             & 3              & 22        & 27          & 9                & 9      & 2           \\
Transit corridors and parking            & 38        & 1      & 2             & 3           & 4       & 4       & 5       & 4               & 0             & 1              & 35        & 22          & 11               & 4      & 1           \\
Public spaces and Parks and recreation   & 22        & 1      & 1             & 3           & 3       & 11      & 11      & 1               & 2             & 0              & 11        & 0           & 0                & 1      & 2           \\
Zoning and rezoning and land use         & 39        & 3      & 0             & 1           & 5       & 1       & 12      & 0               & 1             & 0              & 29        & 0           & 11               & 0      & 0           \\
Local economy                            & 4         & 0      & 0             & 1           & 9       & 5       & 7       & 0               & 0             & 0              & 2         & 0           & 2                & 1      & 0           \\

\midrule
\textit{Social Issues}                   &           &        &               &             &         &         &         &                 &               &                &           &             &                  &        &             \\
Functional democracy                     & 37        & 3      & 12            & 12          & 16      & 86      & 84      & 17              & 9             & 1              & 32        & 5           & 10               & 6      & 0           \\
Affordability                            & 55        & 1      & 0             & 3           & 5       & 41      & 19      & 1               & 3             & 0              & 23        & 0           & 6                & 5      & 0           \\
Public safety                            & 38        & 1      & 0             & 3           & 8       & 26      & 43      & 2               & 3             & 0              & 31        & 1           & 10               & 4      & 2           \\
Quality of the built environment         & 22        & 0      & 1             & 7           & 23      & 11      & 20      & 2               & 1             & 2              & 27        & 16          & 16               & 3      & 0           \\
Homelessness                             & 4         & 1      & 0             & 0           & 1       & 32      & 42      & 0               & 0             & 0              & 3         & 0           & 0                & 2      & 0           \\
Anti-racism                              & 55        & 0      & 0             & 0           & 1       & 29      & 11      & 0               & 0             & 1              & 0         & 0           & 0                & 2      & 0           \\
Sustainability                           & 29        & 0      & 0             & 0           & 2       & 9       & 1       & 2               & 1             & 6              & 20        & 0           & 1                & 7      & 0           \\
Senior/infant/child/and teenager care    & 7         & 0      & 7             & 6           & 22      & 8       & 12      & 3               & 7             & 4              & 12        & 1           & 4                & 2      & 1           \\
Public health                            & 11        & 1      & 3             & 0           & 14      & 15      & 15      & 0               & 3             & 1              & 5         & 0           & 0                & 1      & 1           \\
Incarceration and crime history          & 2         & 0      & 0             & 0           & 0       & 8       & 37      & 0               & 0             & 0              & 0         & 0           & 0                & 0      & 0          
  \\ 
 \bottomrule

\end{tabular}
}
\end{adjustbox}
    \caption{Annotation and prediction by city, numbers refer to comments that contain this local concern or \sc{}}
    \label{tab:anno_city}
\end{table*}

\subsection{Additional Predictive Metrics}

\subsubsection{Model parameters}

See Table \ref{tab:hyperPara} for ranges used for hyperparameter tuning. 

\begin{table*}[hbt!]
    
\begin{tabular}{lll} 
\toprule
Model                                      & Hyperparameters & Range                                        \\ \midrule
\multirow{6}{*}{TF-IDF Vectorizer}         & use\_idf        & \{True\}                                     \\
                                           & lowercase       & \{True\}                                     \\
                                           & stop\_words     & \{None\}                                     \\
                                           & ngram\_range    & \{(1, 1), (1, 2)\}                           \\
                                           & max\_df         & \{0.75, 0.9, 1.0\}                           \\
                                           & min\_df         & \{0.0, 0.1, 0.25\}                           \\  \midrule
\multirow{4}{*}{Logistic Classifier}       & penalty         & \{`l1', `l2', None\}                         \\
                                           & C               & \{1, 10, 100, 1000\}                         \\
                                           & solver          & \{‘lbfgs’, ‘liblinear’, ‘newton-cg’, ‘sag’\} \\
                                           & class\_weight   & \{‘balanced’\}                         \\  \midrule
\multirow{4}{*}{Support Vector Classifier} & C               & \{1, 10, 100, 1000\}                         \\
                                           & gamma           & \{1, 0.1, 0.001, 0.0001\}                    \\
                                           & kernel          & \{‘linear’,‘rbf’\}                           \\ 
                                            & class\_weight   & \{‘balanced’\}                         \\  \midrule
\multirow{5}{*}{DistilBERT, RoBERTa}       & epoch           & \{20\}                                       \\
                                           & batch\_size     & \{16\}                                       \\
                                           & learning\_rate  & \{loguniform(1e-6, 1e-4)\}                   \\
                                           & weight\_decay   & \{loguniform(1e-4, 1e-2)\}                   \\
                                           & dropout         & \{uniform(0.1, 0.5)\} \\ & optimizer         & Adam W  (betas: (0.9, 0.999); epsilon=1e-06)                     \\ 
\bottomrule

\end{tabular}
    \caption{Model hyperparameter tuning was conducted using these ranges to determine the optimal settings.}
    \label{tab:hyperPara}
\end{table*}


\subsubsection{Model results}

See Table \ref{tab:results_long}.

\begin{table*}[hbt!]
    \begin{adjustbox}{width=\textwidth,center}{
\def\sym#1{\ifmmode^{#1}\else\(^{#1}\)\fi}
\begin{threeparttable}
\begin{tabular}{lcc|ccc|ccc|ccc|ccc}
\toprule
      & \multicolumn{2}{l}{\textbf{Annotated Data Overview}} & \multicolumn{3}{c}{\textbf{DistilBERT results}}        & \multicolumn{3}{c}{\textbf{RoBERTa results}}           & \multicolumn{3}{c}{\textbf{Logistics results}}                                                                 & \multicolumn{3}{c}{\textbf{SVC results}}                                                                       \\ 
\textbf{Target Variable}                 & \textbf{Fraction}    & \textbf{Support}     & \textbf{F1}     & \textbf{Precision} & \textbf{Recall} & \textbf{F1}     & \textbf{Precision} & \textbf{Recall} & \multicolumn{1}{c}{\textbf{F1}} & \multicolumn{1}{c}{\textbf{Precision}} & \multicolumn{1}{c}{\textbf{Recall}} & \multicolumn{1}{c}{\textbf{F1}} & \multicolumn{1}{c}{\textbf{Precision}} & \multicolumn{1}{c}{\textbf{Recall}} \\ \midrule
\textsc{\commenttype}       & 87.95\%              & 802                 &   0.7761 & 0.8454 & 0.7458 & 0.8473 & 0.8640 & 0.8353 & \textbf{0.9732} & \textbf{0.9477} & \textbf{1.0000} & 0.9732 & 0.9477 & 1.0000 \\

 & & & (0.0074) & (0.0111) & (0.0074) & (0.0061) & (0.0069) & (0.0085) & (0.0000) & (0.0000) & (0.0000) & (0.0000) & (0.0000) & (0.0000) \\ 

\textsc{Civic}                 & 84.01\% & 767 & 0.7666 & 0.8540 & 0.7535 & 0.7989 & 0.8309 & 0.8032 & \textbf{0.9592} & \textbf{0.9216} & \textbf{1.0000} & 0.9592 & 0.9216 & 1.0000 \\
 &  &  & (0.0100) & (0.0135) & (0.0098) & (0.0092) & (0.0144) & (0.0132) & (0.0000) & (0.0000) & (0.0000) & (0.0000) & (0.0000) & (0.0000) \\

\textsc{Social}              & 78.86\% & 720 & 0.7348 & 0.7610 & 0.7196 & 0.7419 & 0.7759 & 0.7245 & \textbf{0.9333} & \textbf{0.9110} & \textbf{0.9568} & 0.9247 & 0.9214 & 0.9281 \\
 &  &  & (0.00897) & (0.0087) & (0.0090) & (0.0085) & (0.0074) & (0.0091) & (0.0000) & (0.0000) & (0.0000) & (0.0000) & (0.0000) & (0.0000) \\
\midrule
\multicolumn{3}{l}{\textit{Civic areas (average across top 10)}} & \textit{0.7007} & \textit{0.7623} & \textit{0.6851} & \textit{0.7597} & \textit{0.8035} & \textit{0.7499} & \textit{0.3204} & \textit{0.4664} & \textit{0.2994} & \textit{0.1744} & \textit{0.3172} & \textit{0.1662} \\
Public service                              & 21.14\%    & 193   & \textbf{0.5905} & \textbf{0.6333} & \textbf{0.5839} & 0.5585          & 0.6034          & 0.5585          & 0.4037          & 0.5000          & 0.3385          & 0.3738          & 0.4762          & 0.3077          \\
 &  &  & (0.0100) & (0.0155) & (0.0081) & (0.0074) & (0.0070) & (0.0050) & (0.0000) & (0.0000) & (0.0000) & (0.0000) & (0.0000) & (0.0000) \\

Housing                                     & 15.44\%    & 141   & 0.7921          & 0.8293          & 0.7861          & \textbf{0.7921} & \textbf{0.8442} & \textbf{0.7799} & 0.6000          & 0.7500          & 0.5000          & 0.4828          & 0.8750          & 0.3333          \\
&  &  & (0.0050) & (0.0095) & (0.0070) & (0.0096) & (0.0099) & (0.0107) & (0.0000) & (0.0000) & (0.0000) & (0.0000) & (0.0000) & (0.0000) \\
Election administration and Appointments    & 11.28\%    & 103   & 0.6822          & 0.8818          & 0.6416          & \textbf{0.7238} & \textbf{0.8475} & \textbf{0.6868} & 0.3784          & 0.7000          & 0.2593          & 0.0000          & 0.0000          & 0.0000          \\
 &  &  & (0.0102) & (0.0133) & (0.0086) & (0.0096) & (0.0167) & (0.0084) & (0.0000) & (0.0000) & (0.0000) & (0.0000) & (0.0000) & (0.0000) \\
Policing                                    & 8.43\%     & 77    & 0.8095          & 0.8658          & 0.7787          & \textbf{0.8652} & \textbf{0.8859} & \textbf{0.8499} & 0.6207          & 0.6923          & 0.5625          & 0.5385          & 0.7000          & 0.4375          \\
 &  &  & (0.0097) & (0.0057) & (0.0132) & (0.0065) & (0.0103) & (0.0056) & (0.0000) & (0.0000) & (0.0000) & (0.0000) & (0.0000) & (0.0000) \\
Utility service                             & 7.01\%     & 64    & 0.7320           & 0.7857          & 0.7207          & \textbf{0.7993} & \textbf{0.8092} & \textbf{0.7965} & 0.0000          & 0.0000          & 0.0000          & 0.0000          & 0.0000          & 0.0000          \\
 &  &  & (0.0105) & (0.0173) & (0.0125) & (0.0104) & (0.0082) & (0.0139) & (0.0000) & (0.0000) & (0.0000) & (0.0000) & (0.0000) & (0.0000) \\
Public works                                & 6.90\%     & 63    & 0.6029          & 0.6688          & 0.5892          & \textbf{0.6915} & \textbf{0.6966} & \textbf{0.6895} & 0.1667          & 0.2500          & 0.1250          & 0.1951          & 0.1212          & 0.5000          \\
 &  &  & (0.0104) & (0.0152) & (0.0087) & (0.0139) & (0.0161) & (0.0126) &(0.0000) & (0.0000) & (0.0000) & (0.0000) & (0.0000) & (0.0000) \\
 
Zoning and rezoning and land use            & 6.68\%     & 61    & 0.7627          & 0.8108          & 0.7367          & \textbf{0.7998} & \textbf{0.8489} & \textbf{0.7675} & 0.1538          & 1.0000          & 0.0833          & 0.1538          & 1.0000          & 0.0833          \\
 &  &  & (0.0096) & (0.0162) & (0.0076) & (0.0087) & (0.0135) & (0.0094) & (0.0000) & (0.0000) & (0.0000) & (0.0000) & (0.0000) & (0.0000) \\

Transit corridors and parking               & 6.24\%     & 57    & 0.7772          & 0.7715          & 0.8010           & \textbf{0.8543} & \textbf{0.8239} & \textbf{0.9036} & 0.7273          & 0.5714          & 1.0000          & 0.0000          & 0.0000          & 0.0000          \\ 
 &  &  & (0.0104) & (0.0162) & (0.0095) & (0.0099) & (0.0110) & (0.0132) & (0.0000) & (0.0000) & (0.0000) & (0.0000) & (0.0000) & (0.0000) \\

Public spaces and Parks and recreation      & 5.70\%     & 52    & 0.7362          & 0.8257          & 0.6938          & \textbf{0.9375} & \textbf{0.9898} & \textbf{0.9149} & 0.1538          & 0.2000          & 0.1250          & 0.0000          & 0.0000          & 0.0000          \\
 &  &  & (0.0166) & (0.0225) & (0.0154) & (0.0167) & (0.0144) & (0.0196) & (0.0000) & (0.0000) & (0.0000) & (0.0000) & (0.0000) & (0.0000) \\

Local economy                               & 2.85\%     & 26    & 0.5216          & 0.5509          & 0.5189          & \textbf{0.5744} & \textbf{0.6853} & \textbf{0.5520}  & 0.0000          & 0.0000          & 0.0000          & 0.0000          & 0.0000          & 0.0000         \\
 &  &  & (0.0217) & (0.0459) & (0.0146) & (0.0254) & (0.0612) & (0.0166) & (0.0000) & (0.0000) & (0.0000) & (0.0000) & (0.0000) & (0.0000) \\

 \midrule
\multicolumn{3}{l}{\textit{Social Issues (average across top 10)}} & \textit{0.7870} & \textit{0.8270} & \textit{0.7761} & \textit{0.8130} & \textit{0.8352} & \textit{0.8089} & \textit{0.4908} & \textit{0.5972} & \textit{0.5188} & \textit{0.2847} & \textit{0.5041} & \textit{0.2005} \\
Functional democracy                        & 27.38\%     & 250    & 0.7032          & 0.7234          & 0.6956          & \textbf{0.7403} & \textbf{0.7495} & \textbf{0.7370} & 0.6963          & 0.8103          & 0.6104          & 0.5487          & 0.8611          & 0.4026          \\
 &  &  & (0.0037) & (0.0046) & (0.0046) & (0.0064) & (0.0064) & (0.0077) & (0.0000) & (0.0000) & (0.0000) & (0.0000) & (0.0000) & (0.0000) \\
 
Affordability                               & 13.58\%     & 124    & 0.7675          & 0.7966          & 0.7596          & \textbf{0.8380} & \textbf{0.8620} & \textbf{0.8225} & 0.4552          & 0.6316          & 0.3571          & 0.5263          & 1.0000          & 0.3571          \\
 &  &  & (0.0102) & (0.0157) & (0.0088) & (0.0099) & (0.0129) & (0.0112) & (0.0000) & (0.0000) & (0.0000) & (0.0000) & (0.0000) & (0.0000) \\

Public safety                               & 13.03\%     & 119    & 0.8021          & 0.8126          & 0.8034          & \textbf{0.8378} & \textbf{0.8286} & \textbf{0.8511} & 0.3824          & 0.2889          & 0.5652          & 0.4118          & 0.6364          & 0.3043          \\
 &  &  & (0.0105) & (0.0139) & (0.0144) & (0.0059) & (0.0083) & (0.0049) & (0.0000) & (0.0000) & (0.0000) & (0.0000) & (0.0000) & (0.0000) \\

Anti-racism                                 & 10.51\%     & 96     & \textbf{0.8220} & \textbf{0.8146} & \textbf{0.8507} & 0.7914          & 0.7717          & 0.8269          & 0.1818          & 0.3333          & 0.1250          & 0.0000          & 0.0000          & 0.0000          \\
 &  &  & (0.0083) & (0.0085) & (0.0114) & (0.0087) & (0.0078) & (0.0133) & (0.0000) & (0.0000) & (0.0000) & (0.0000) & (0.0000) & (0.0000) \\

Quality of the built environment            & 9.20\%      & 84     & 0.6484          & 0.7089          & 0.6277          & \textbf{0.7169} & \textbf{0.7740} & \textbf{0.6863} & 0.1667          & 0.2500          & 0.1250          & 0.0690          & 0.0769          & 0.0625          \\
 &  &  & (0.0198) & (0.0218) & (0.0213) & (0.0159) & (0.0219) & (0.0134) &(0.0000) & (0.0000) & (0.0000) & (0.0000) & (0.0000) & (0.0000) \\

Homelessness                                & 8.76\%      & 80     & 0.8851          & 0.9488          & 0.8581          & \textbf{0.9028} & \textbf{0.9369} & \textbf{0.8760} & 0.7353          & 0.6579          & 0.8333          & 0.4000          & 0.8000          & 0.2667          \\ 
 &  &  & (0.0070) & (0.0068) & (0.0116) & (0.0044) & (0.0036) & (0.0067) & (0.0000) & (0.0000) & (0.0000) & (0.0000) & (0.0000) & (0.0000) \\

Senior,infant, child, and teenager care     & 6.79\%      & 62     & 0.7390          & 0.8289          & 0.7196          & \textbf{0.7662} & \textbf{0.8569} & \textbf{0.7453} & 0.2222          & 1.0000          & 0.1250          & 0.0000          & 0.0000          & 0.0000          \\
 &  &  & (0.0113) & (0.0130) & (0.0170) & (0.0087) & (0.0172) & (0.0086) & (0.0000) & (0.0000) & (0.0000) & (0.0000) & (0.0000) & (0.0000) \\

Public health                               & 6.46\%      & 59     & 0.7918          & 0.8113          & 0.7882          & \textbf{0.8047} & \textbf{0.7887} & \textbf{0.8323} & 0.6154          & 0.5000          & 0.8000          & 0.3077          & 0.6667          & 0.2000          \\
 &  &  & (0.0139) & (0.0163) & (0.0201) & (0.0133) & (0.0104) & (0.0191) &(0.0000) & (0.0000) & (0.0000) & (0.0000) & (0.0000) & (0.0000) \\

Incarceration and crime history             & 5.15\%      & 47     & 0.8987          & 0.9594          & 0.8595          & \textbf{0.8996} & \textbf{0.9466} & \textbf{0.8720} & 0.7857          & 1.0000          & 0.6471          & 0.5833          & 1.0000          & 0.4118          \\
 &  &  & (0.0081) & (0.0112) & (0.0095) & (0.0055) & (0.0123) & (0.0007) & (0.0000) & (0.0000) & (0.0000) & (0.0000) & (0.0000) & (0.0000) \\

Sustainability                              & 4.49\%      & 41     & 0.8118          & 0.8653          & 0.7989          & \textbf{0.8327} & \textbf{0.8369} & \textbf{0.8399} & 0.6667          & 0.5000          & 1.0000          & 0.0000          & 0.0000          & 0.0000         \\ 
 &  &  & (0.0136) & (0.0193) & (0.0136) & (0.0147) & (0.0095) & (0.0225) & (0.0000) & (0.0000) & (0.0000) & (0.0000) & (0.0000) & (0.0000) \\

\bottomrule
\end{tabular}
\begin{tablenotes}
    \item Results are averaged across 3 random seeds.
    \item Standard errors are shown in parenthesis.
    \item Best models (models with the highest F1) are emphasized in bold.
 
\end{tablenotes}
\end{threeparttable}
}
\end{adjustbox}
    \caption{Predictive results on test data (1/6 of annotated data)}
    \label{tab:results_long}
\end{table*}

\begin{algorithm}[ht]
\caption{Gold Set Selection Algorithm for selecting cities for which to collect data}
\label{alg:goldset}
\begin{algorithmic}[1]
\Require A set of municipalities \( \mathcal{M} \) with demographic and political attributes, 
         and the corresponding state-level probability vectors.
\Ensure A subset of municipalities \( \mathcal{G} \) (the “Gold Set”) that minimizes 
        divergence from the state-level distribution.

\State Initialize \(\mathcal{G} \leftarrow \varnothing\).
\State Let \(N\) be the number of random subsets to sample (e.g., thousands of subsets).
\For{each random subset \( \mathcal{S}_i \subseteq \mathcal{M} \), for \(i = 1 \ldots N\)}
    \State Construct probability vectors \( \mathbf{p}_{\text{pop}}(\mathcal{S}_i) \) and \( \mathbf{p}_{\text{pol}}(\mathcal{S}_i) \) 
           for population and political leaning within \( \mathcal{S}_i \).
    \State Compute the Kullback-Leibler (KL) divergences:
          \[
            D_{\mathrm{KL}}^{(\text{pop})} = D_{\mathrm{KL}}\big(\mathbf{p}_{\text{pop}}(\mathcal{S}_i) \parallel \mathbf{p}_{\text{pop}}(\text{Michigan})\big),
          \]
          \[
            D_{\mathrm{KL}}^{(\text{pol})} = D_{\mathrm{KL}}\big(\mathbf{p}_{\text{pol}}(\mathcal{S}_i) \parallel \mathbf{p}_{\text{pol}}(\text{Michigan})\big).
          \]
    \State Compute the average divergence for \( \mathcal{S}_i \):
          \[
            \bar{D}_{\mathrm{KL}}( \mathcal{S}_i ) = \frac{D_{\mathrm{KL}}^{(\text{pop})} + D_{\mathrm{KL}}^{(\text{pol})}}{2}.
          \]
\EndFor
\State Identify the subset \(\mathcal{S}^* \) with the smallest \(\bar{D}_{\mathrm{KL}}\) value:
      \[
        \mathcal{S}^* = \arg\min_{\mathcal{S}_i} \big(\bar{D}_{\mathrm{KL}}( \mathcal{S}_i )\big).
      \]
\State Let \(\mathcal{G} \leftarrow \mathcal{S}^*\) be the selected Gold Set.
\State \textbf{return} \(\mathcal{G}\)

\end{algorithmic}
\end{algorithm}

\end{document}